\documentclass{jfm}

\usepackage{graphicx}
\usepackage{newtxtext}
\usepackage{newtxmath}
\usepackage{natbib}
\usepackage{placeins}
\usepackage{hyperref}
\usepackage{amssymb}
\usepackage{dsfont}
\usepackage[justification=justified,width=\textwidth]{caption}
\hypersetup{
    colorlinks = true,
    urlcolor   = blue,
    citecolor  = black,
}

\definecolor{orange}{RGB}{221,110,78}

\newcommand{\RomanNumeralCaps}[1]
\linenumbers

\title{Morphology of ice structures induced by a freezing rivulet}

\author{Hélie de Miramon\aff{1}
  \corresp{\email{heliedemiramon@gmail.com}},
  Wladimir Sarlin\aff{2,3}, Christophe Josserand\aff{1}, Thomas Séon\aff{3} \and Axel Huerre\aff{2}}

\affiliation{\aff{1}Laboratoire d’Hydrodynamique (LadHyX), UMR 7646 CNRS-Ecole Polytechnique, IP Paris, 91128 Palaiseau CEDEX, France
\aff{2}Laboratoire Matière et Systèmes Complexes (MSC), Université Paris Cité, CNRS, UMR 7057, Paris, 75013, France
\aff{3}Institut Franco-Argentin de Dynamique des Fluides pour l’Environnement (IFADyFE), CNRS (IRL 2027), Universidad de Buenos Aires, CONICET, Buenos Aires, Argentina
}

\begin{document}
\maketitle

\begin{abstract}

We investigate the solidification of a water rivulet flowing over a cold inclined substrate and the resulting formation of three-dimensional ice structures. Using a controlled hydraulic and thermal setup, combined with spatiotemporal phase-shifting profilometry and infrared thermography, we characterize both the transient evolution and the final morphology of the ice. We show that a typical experiment proceeds through three stages: formation of a straight ice ridge that stabilizes the rivulet, destabilization and lateral excursions of the flow leading to rapid transverse spreading of the ice structure, and progressive thickening and smoothing of the ice block. Across a wide range of flow rates, inclinations and thermal conditions, the final morphology comprises an upstream triangular lateral envelope, followed by a downstream region of nearly constant width once the substrate edges are reached. Infrared measurements reveal that the rivulet residence time on the substrate follows a Gaussian distribution in the azimuthal angle, implying that the central region of the structure is visited far more frequently than its lateral edges. Focusing on the domain where the height has converged at the end of the experiments, we develop a two-dimensional theoretical model that couples a hydrodynamic model for the rivulet geometry with heat transport in both liquid and solid phases. In the large Péclet number limit, the model predicts an exponential increase of the stationary ice height along the flow direction and it shows an excellent agreement with the experimental height field. We further show how the combination of the stationary height profile and the non-uniform residence time distribution controls the angular convergence of the ice cross-sections.

\end{abstract}

\begin{keywords}
Authors should not enter keywords on the manuscript, as these must be chosen by the author during the online submission process and will then be added during the typesetting process (see \href{https://www.cambridge.org/core/journals/journal-of-fluid-mechanics/information/list-of-keywords}{Keyword PDF} for the full list).  Other classifications will be added at the same time.
\end{keywords}

\section{Introduction}

The Earth, and particularly its coldest regions, displays numerous examples in which natural flows interact with phase change to generate complex ice morphologies. In underground caves, for instance, thin rivulets or films of water can freeze onto overhanging cold surfaces, forming ice draperies \citep{Persoiu_2017}, as well as ice stalactites and stalagmites growing from dripping water \citep{Chen_2011, Papa_2025}, while at larger scales similar processes produce ice waterfalls along steep mountain faces \citep{Montagnat_2010}. On glacier surfaces, the interaction between flowing meltwater and the melting substrate gives rise to a large variety of morphologies collectively referred to as glacial cryokarst. During the melt season, surface meltwater accumulates in depressions to form supraglacial lakes and organizes into streams and rivers that incise the ice \citep{Pitcher_2019}. These flows may eventually plunge into crevasses, eventually leading to the formation of englacial conduits at the glacier base \citep{Mankoff_2017}. By modulating subglacial hydrology, such drainage systems exert a strong influence on glacier motion and stability. With glaciers having lost nearly 5\% of their volume between 2000 and 2023 \citep{Zemp_2025}, it has become increasingly important to determine how supraglacial streams may accelerate the melting of the very surfaces they flow over \citep{Pitcher_2019}. Even though the formation, evolution and dynamics of supraglacial streams has gained considerable attention among the scientific community in recent years \citep{Pitcher_2019}, their approach mostly consists of field work (as soon as the early 20th century through the expedition of David Shackleton \citep{David_1910}) and more recently remote sensor analysis \citep{Smith_2015, Yang_2016, St_Germain_2019}. Despite their crucial role in glacier mass loss and dynamics, supraglacial rivers remain poorly understood. Indeed, their global distribution is only partly mapped due to the logistical difficulty of acquiring field observations in remote regions \citep{Gleason_2016}, while their incision rate and rapid morphological adjustments still lack quantitative validation at field scale \citep{Karlstrom_2013}. Moreover, current ice-sheet models do not yet represent the routing of meltwater through channelized flow paths or the delivery of these surface discharges to the glacier base, despite growing evidence that such processes strongly modulate ice dynamics and ice-shelf stability \citep{Kingslake_2017}, leaving major uncertainties in predictions of glacier stability and sea-level rise \citep{Pitcher_2019}.

Yet, beyond these gravity-dominated flows, similar interactions between a moving liquid and an evolving ice surface also occur at much smaller, capillary scales \citep{Huerre_2025}. Studying such configurations offers a complementary route to isolate the fundamental physical mechanisms governing flow–ice coupling and apply them to understand the formation of larger scale ice structures. This is typically exemplified by the accretion of ice on man-made structures \citep{Lynch_2001, Baumert_2018, Liu_2019} resulting from the repeated solidification of impacting water droplets. Several works have therefore addressed the impact of droplets on cold substrates. \citet{Madejski_1976} conducted an experimental and theoretical study describing the spreading dynamics of a droplet impacting a substrate whose surface temperature lies below the liquid’s melting point, basing his analysis on energetic arguments. \citet{Ghabache_2016} investigated fracture regimes that develop in the ice formed by the solidification of an impacting droplet. \citet{Schremb_2018} characterized the dynamics of the liquid film following the impact of a supercooled water droplet on ice, as well as the resulting final ice shape. The maximum spreading diameter reached upon droplet impact on a cold substrate was studied by \citet{Thievenaz_2020b}, who proposed an effective-viscosity model to account for their experimental observations. This model was later extended to the case in which the droplet is capable of melting its substrate during spreading \citep{Sarlin_2024}, thereby capturing behaviors reported in previous studies \citep{Jin_2017, Ju_2019}. \citet{Papa_2025} also explored the case of repeated droplet impacts on a cold substrate, which leads to the formation of stalagmites of frozen liquid. The spreading dynamics of a water droplet deposited on ice were recently examined by \citet{Sarlin_2025}, who determined the macroscopic contact angle of water on ice. It should be mentioned, also, that numerous studies have addressed the morphology of the ``Byzantine frozen drop", i.e. the ice shape resulting from the solidification of a sessile droplet, although this problem remains open \citep{Stairs_1971, Sebilleau_2021}.

Droplets, however, are not the only capillary flows of interest. When a fluid flows down an inclined plane, a variety of hydrodynamic regimes are observed depending on the volumetric flow rate $Q$ \citep{Nakagawa_1984, Nakagawa_1992}. As $Q$ increases, one may indeed observe dripping, straight rivulets \citep{Monier_2022}, meanders \citep{Piteira_2006}, or films exhibiting a braided pattern \citep{Mertens_2004}. The solidification of a straight rivulet was recently investigated by \citet{Monier_2020}, who described its dynamics and reported the formation of a surprising ice morphology whose height increases linearly with the distance from the injection point. This shape, along with the temperature fields in the various phases of the problem, was subsequently captured by an advection–diffusion model coupled to a Stefan condition at the water–ice interface \citep{Huerre_2021}. Yet, these studies focused on short timescales and on flow conditions specifically chosen to ensure the stability of a straight rivulet. Other rivulet regimes (in particular at higher flow rates where lateral instabilities and branching occur) remain largely unexplored, even in the absence of solidification \citep{Piteira_2006, Aouad_2016}. Extending solidification experiments to these more complex hydrodynamic regimes is therefore essential to uncover the range of ice morphologies that may develop, and to clarify the physical mechanisms responsible for their formation in natural and engineered environments.

The present manuscript aims to provide elements of response to these latter questions. It is organized as follows. We first introduce the experimental setup and methods and then we will describe the evolution and dynamics of a typical experiment, providing general explanation of the emergence of three dimensional ice structures in our range of control parameters. Both the dynamic spreading and the convergence of the structure towards a stationary state will be analyzed. Next, we develop a two-dimensional theoretical model for the stationary ice height field, coupling the rivulet hydrodynamics with heat transport in both the liquid and the solid domains. The model predictions are compared successuflly with experimental measurements. Finally, we show that the angular dependence of the ice structures arises from the Gaussian distribution of the rivulet residence time on the cold substrate, as it spends far more time on its centerline compared to its edges.

\section{Experimental methods and setup}

\begin{figure}
    \centering
    \includegraphics[width=\linewidth]{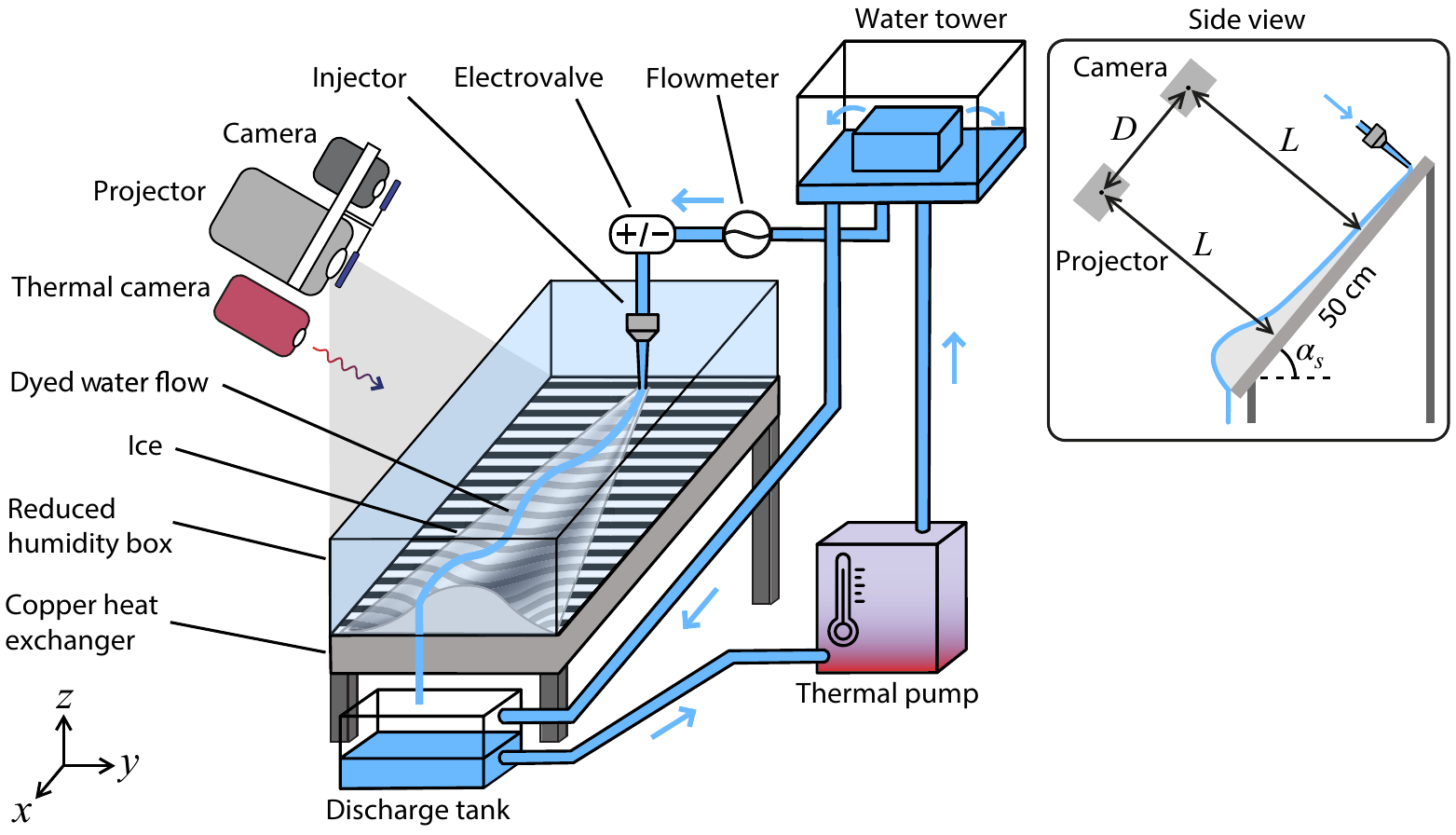}
    \caption{Schematic of the experimental setup used in this study. It consists of a hydraulic loop providing a constant water flow injected at the top of an inclined substrate, a cryostat-regulated copper heat exchanger ensuring controlled surface temperature, and an optical system (consisting of a projector/camera pair) that enables three-dimensional mapping of the ice structure height field using the Spatiotemporal Phase-Shifting Profilometry method. A sealed low-humidity enclosure prevents frost formation throughout the 18h experiments.}
    \label{Dispositif experimental}
\end{figure}

The experimental setup schematized in Figure~\ref{Dispositif experimental} is used to study the formation of the ice structure induced by the freezing of a rivulet on a cold substrate. It is made up of three distinct parts: a hydraulic system that ensures a constant flow rate throughout the duration of the experiment ($\mathrm{18 h}$), a thermal system to control the temperature at the surface of the substrate and an optical setup that enables three-dimensional reconstructions of the ice structure over time using an optical profilometry technique. The design allows us to consider that all control parameters are constant quasi-instantaneously, and for the total duration of an experiment (18h), apart for the injected water temperature that reaches a final value of interest after a transition period smaller than 10 minutes. The following paragraphs detail the three subsystems independently.

\subsection{Flow-rate control}

The hydraulic setup is a closed loop that allows us to maintain a constant flow rate $Q$ of water throughout an experiment. Here, $Q$ was varied between 15 and 200 $\mathrm{mL.min^{-1}}$ in order to cover the transition from straight rivulet flow to meandering behavior typically observed on a copper substrate at ambient temperature. As illustrated in the schematics in Figure~\ref{Dispositif experimental}, the heat exchanger is mounted on top of a discharge water tank and features a rotating axis at the top, allowing us to vary the substrate inclination angle $\alpha_s$. In this study, three configurations have been considered,  $\alpha_s \in \{50.3^{\circ}, 90^{\circ}, 139.6^{\circ}\}$.

The water is pumped from the discharge tank to the water tower located approximately 50 cm above the heat exchanger. The flow rate is monitored using an electromagnetic flowmeter (Kobold MIM) and controlled with an electro-valve. Both the electro-valve and the flowmeter are connected to a controller (RB 100), enabling the feedback loop to maintain the target flow rate constant throughout the experiment. Water is then injected through a metallic needle with an inner diameter $d_{in}=3 \ \rm{mm}$, using an injector placed 4 mm above the substrate (perpendicularly to the incline), to prevent flow destabilization and droplet formation at lower flow rates.

All experiments begin with the the opening of the electro valve at $t=0$, reaching the flow rate of interest after only a few seconds, without any overshoot. To avoid later disturbances in flow rate regulation, the electro valve is then controlled with an integration time of 5 seconds, filtering out the instantaneous fluctuations inherent to the flowmeter. This ensures a stable injected flow rate throughout the 18 hours of study, with typical standard deviations ranging from $0.8 \ \rm{mL.min^{-1}}$ for $Q=15 \ \rm{mL.min^{-1}}$, to $2-3 \ \rm{mL.min^{-1}}$ for $Q=200 \ \rm{mL.min^{-1}}$. 

\subsection{Temperatures control}

In order to have a surface of constant temperature even with a water flow on top of it, a copper heat exchanger of size 50x20x2 $\rm{cm}^3$ is connected to a cryostat (Lauda RP290E). Specific heat transfer liquid (Lauda Kryo 95 oil) flows from the cryostat into the heat exchanger, thus enabling a spatially homogeneous and heat transfer resistant surface at any temperature ranging from -60 to 100°C. Temperature regulation is monitored using three Class 1/3 PT100 probes (accuracy $\pm 0.1^\circ\mathrm{C}$) inserted on the side of the heat exchanger at mid-thickness: two positioned near the upstream and downstream ends of the plate to measure its temperature distribution, and a third located midway along its length to provide feedback control to the cryostat. Due to the high cooling capacity of the thermal pump ($640-800 \, \mathrm W$ for desired temperatures ranging from $-40$ to $20^{\circ} \rm{C}$), the surface temperature is kept constant with a typical standard deviation $\sigma (T_s) < 0.4^{\circ}\rm{C}$. Across the experiments presented in this study, $T_s$ spans $-40^\circ \mathrm C$ to $-5^\circ \mathrm C$.

The water temperature is controlled using a Huber Minichiller 600, and is initially cooled before being pumped to the water tower. Due to the ambient air heating up the dyed water flowing inside the pipes, the injected water temperature typically transitions to the temperature of study after 1-10 minutes. After this transition, the temperature of the injected water remains stable throughout the experiment, with typical standard deviation $\sigma(T_\mathrm{in}) < 0.5 ^{\circ} \rm{C}$. Across the experiments presented in this study, $T_\mathrm{in}$ spans $4^\circ \mathrm C$ to $34^\circ \mathrm C$.

Finally, to prevent the formation of frost, a transparent 45x52x24 $\rm{cm}^3$ airtight plexiglass box is added on top of the heat exchanger and pressurized dry air is injected inside it at a flow rate $Q_{\rm air}=200 \ \rm{L.min^{-1}}$ ensuring a reduced humidity of around $8-14\%$ for all experiments. 

\subsection{Optical setup}

\subsubsection{Principle of the profilometry method}

Among the various existing profilometry methods, the chosen method in this study is called Spatio-Temporal Phase-Shifting Profilometry (ST-PSP). This profilometry method is based on the projection of a sinusoidal fringe pattern onto a reference plane (here, the substrate) using a video projector. When an object is present, the fringes get deformed and this deformation can be recorded using a camera, as schematized in Figure~\ref{Dispositif experimental}. The analysis of the obtained images allows us to determine their phase shift with the reference plane, and consequently determine the height map of the object \citep{Maurel_2009, de_Miramon_2025}. 

The ST-PSP method combines both the Sampling Moiré and Phase-Shifting Profilometry methods to improve the accuracy and robustness of phase determination \citep{Ri_2019}. This is achieved by processing a set of $N$ phase-shifted images and performing down-sampling and interpolation across all $N$ images. The phases $\phi$ and $\phi_0$ of the object and the reference plane can then be inferred by a two-dimensional discrete Fourier transform \citep{de_Miramon_2025}. One can then link the height map of the object with the phase difference $\Delta \phi=\phi-\phi_0$ between the object and the reference plane (extracted separately with the aforementioned methodology). This link was first established by \citet{Takeda_1983} and was then refined by \citet{Maurel_2009}. Here, the non-collimated projection with parallel-optical-axes geometry is used, so that the phase-to-height relation is \citep{Maurel_2009}:

\begin{equation}
    \label{eq_PTH_exact}
    h(x',y') = \frac{L\Delta \phi(x,y)}{\Delta \phi(x,y) - \omega_0 D},
\end{equation}

where $D$ is the perpendicular distance between the projector and the camera optical axes, $L$ is the distance between the position of the entrance pupils and the reference plane and $\omega_0$ is the frequency of the projected fringes (see the inset of Figure~\ref{Dispositif experimental}). One can observe a shift in coordinates between both sides of Equation~\eqref{eq_PTH_exact}, due to parallax effects. This shift is defined by

\begin{equation}
    \label{xy_prime}
    x^\prime = \left( 1 - \frac{h(x^\prime,y^\prime)}{L} \right)x,
    \qquad
    y^\prime = \left( 1 - \frac{h(x^\prime,y^\prime)}{L} \right)y.
\end{equation}

When the height of the measured object is small compared with the projection distance $L$, parallax effects can be neglected and the phase-to-height relation can be applied using the same coordinates on both sides of Equation~\eqref{eq_PTH_exact}. In our configuration, however, the ice structure can reach heights ranging from 5 to 10\% of $L$, so parallax cannot be ignored. We therefore account for these effects by solving Equation~\eqref{eq_PTH_exact} implicitly, through regridding and interpolation of the height field from the non-linear $(x', y')$ coordinates to the regular $(x, y)$ grid \citep{Grivel_2018}.

\subsubsection{Implementation of the ST-PSP method}

A white aqueous liquid dye containing titanium dioxide and chalk particles (Edding 4090 marker, $C=1 \ \rm{g.L^{-1}}$), is added to the water solution to enhance contrast for the 3D ice profile \citep{de_Miramon_2025}. The experimental setup is completed by a projector and a camera to provide three dimensional reconstructions using the ST-PSP method. An Epson TW7100 projector (4K definition) is used, alongside with a Nikon D810 camera with a $50~\rm{mm}$ Nikon lens, resulting in 36MP images. Both the camera and projector are positioned perpendicular to the substrate at a distance $L \simeq 1.32\ \rm{m}$ and are separated by a distance $D \simeq 28\ \rm{cm}$, as illustrated in the schematic side view in Figure~\ref{Dispositif experimental}. To eliminate parasitic reflections on the surface of interest, two crossed linear polarizers (100mm SQ TS, Edmund Optics) were added to the camera and the projector. All measurements reported in this study were performed using the minimum number of phase-shifted fringe patterns, i.e. $N = 3$, so as to minimize acquisition time for each measurement \citep{Zuo_2018}. The photos were taken with a time interval of $\Delta t \simeq 3.3\ \rm{s}$, which yields approximately one reconstruction every 10 seconds. Given the long duration of the experiments and the resulting volume of data, the acquisition interval is gradually increased: from 30 seconds after the first hour, and then to 2 minutes after four hours.

\subsection{IR measurements}

For some experiments where the position of the water flow on top of the ice structure is investigated, a thermal camera (FLIR A600-Series) is used and is positioned perpendicular to the substrate, so that the surface temperature field can be recorded. This, in turn, enables continuous tracking of the rivulet’s position by distinguishing between positive and negative temperature regions. To enable infrared measurements with the thermal camera, the upper face of the low-humidity box (made of Plexiglas which is opaque to IR radiation) was removed for these particular experiments.

\section{General description of the ice structure formation}

When a water rivulet flows over a cold substrate, its solidification progressively shapes the underlying ice. This mutual interaction (the flow sculpting the ice while the evolving ice redirects the flow) drives the emergence of complex morphologies. In the following, we describe this formation process, from the initial straight ridge produced upon first contact with the cold substrate to the three-dimensional geometry observed at the end of the experiments. Due to the precise control parameter regulation, the experiments reported here are highly reproducible under identical operating conditions. To illustrate the typical dynamics of the ice structure formation, we will focus in this section on a "reference experiment" performed at $\alpha_s^\mathrm{ref}=50.3^\circ$, $Q^\mathrm{ref}=67\ \mathrm{mL,min^{-1}}$, $T_\mathrm{in}^\mathrm{ref}=19.9^\circ\mathrm{C}$ and $T_s^\mathrm{ref}=-22.1^\circ\mathrm{C}$. The behaviors highlighted below are representative of all experiments conducted within the range of conditions explored in this study.

\subsection{Overall evolution of the ice structure}

\begin{figure}
    \centering
    \includegraphics[width=\linewidth]{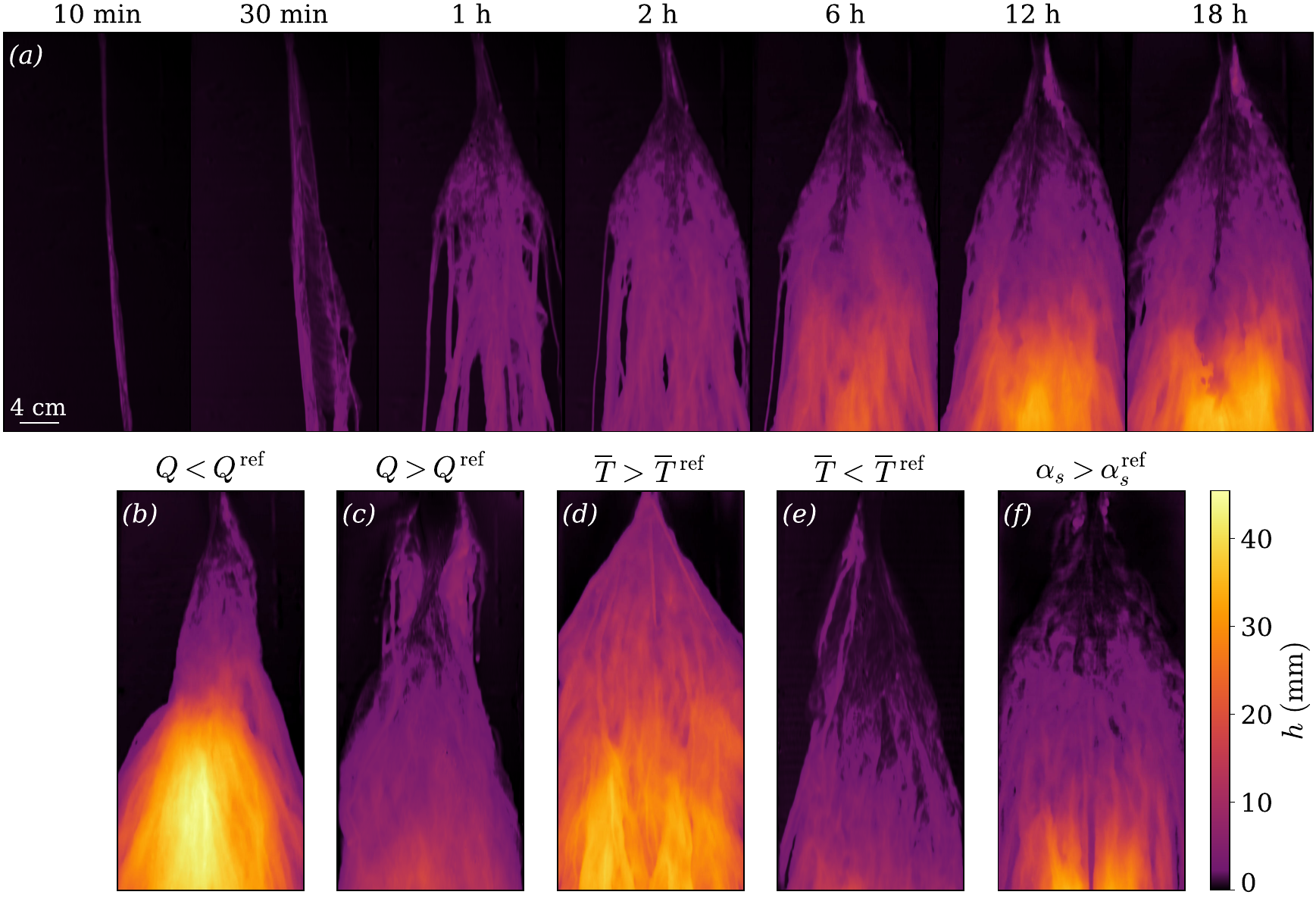}
    \caption{(a) Evolution of the reconstructed ice structure during the reference experiment ($\alpha_s^\mathrm{ref}=50.3^\circ$, $Q^\mathrm{ref}=67 \ \rm{mL.min^{-1}}$, $\overline{T}^\mathrm{ref}=1.11$). Water flows from top to bottom. The final state of the reference experiment is then compared to 5 different experiments (b-f), with each corresponding to a variation of a single control parameter, summarized at the top of each image: (b) $Q=30 \ \mathrm{mL.min^{-1}}$, (c) $Q=193 \ \mathrm{mL.min^{-1}}$, (d) $\overline{T}=0.48$, (e) $\overline{T}=3.88$ and (f) $\alpha_s=90^\circ$. A single color scale, shown at the bottom right of the Figure, is used for all reconstructions.}
    \label{Macrostructure et differents etats finaux}
\end{figure}

To characterize the dynamics of this reference experiment, Figure~\ref{Macrostructure et differents etats finaux}(a) presents the temporal evolution of the ice structure from its initiation to its final state. All images correspond to three dimensional reconstructions of the ice structure at different times of the experiment, from the initial straight rivulet at $t \sim 10 \ \rm{min}$ to the ice block obtained after $t = 18 \rm{h} $. The object's height field is encoded by the colorbar located at the bottom right of the Figure.

In the beginning, at $t\leq 10 \, \mathrm{min}$, a quasi-straight ice ridge emerges as soon as the rivulet starts to run down the exchanger. It should be noted that a straight rivulet with the same flow rate flowing down a copper incline at ambient temperature would instantaneously destabilize and eventually form meanders, as the experimental flow rate is higher than the critical flow rate characterizing the appearance of meanders \citep{Piteira_2006}. However, when flowing down a cold incline at sub-zero temperatures, the solidification process exerts a stabilizing influence on the flow, disturbing therefore the inertia-capillary balance \citep{de_Ruiter_2017}. As the rivulet flows over the ice ridge, the solidified ice locally modifies the wetting properties compared to those of the copper surface, notably by increasing the apparent contact angle \citep{Monier_2022}. The ice thus acts as a surface defect, significantly enhancing contact-angle hysteresis. As a result, for the rivulet to move laterally and initiate a meandering motion, it would need a large depinning force. This mechanism explains the remarkable stability of the rivulet flowing atop the ice ridge during the early stage of the experiment, until about 30 min. After this period of continuous thickening of the initial ice ridge, the rivulet inevitably destabilizes on the bottom half of the exchanger, spilling over the sides of the ice wall, thereby inducing an increase of the area covered by ice, as illustrated by the image for $t=30 \ \mathrm{min}$. It can either amplify continuously, spreading along the transverse axis, or lead to the formation of new ice branches, with one typical example illustrated on the left side of the reconstruction at $t=1 \rm{h}$. A more detailed analysis of the rivulet’s trajectory during the first hour of the experiment will be provided in Section~\ref{sec spreading and rivulet oscillations}.

Following this initial destabilization, the rivulet slowly shifts laterally, creating new branches and moving sideways ($t=2\rm{h}$). During this phase, the ice structure gradually thickens, progressively filling the gaps between neighboring branches ($t=6\rm{h}$). When viewed from above, the overall shape of the ice structure evolves into a quasi-triangular shape as the surface smoothens and the individual branches coalesce into a single, continuous ice block (from $t=6 \rm{h}$ to $t=18 \rm{h}$). It is also observed that the ice layer thickens with increasing distance from the injector: as the flow moves farther away on the ice structure, it becomes colder, which reduces the heat flux delivered by the rivulet to the growing ice structure, accelerates the advance of the solidification front, and consequently increases the ice thickness. During the last hours of the experiment ($t=12 \rm{h}$ to $t=18 \rm{h}$), the area covered by ice ceases to extend laterally, but continues to thicken slowly, particularly near its lower end. 

Following the analysis of the temporal evolution of the ice structure for this reference case, it is useful to analyze how the final ice morphology depends on the control parameters: the substrate inclination $\alpha_s$, the rivulet flow rate $Q$, the injection temperature $T_\mathrm{in}$, and the substrate temperature $T_s$. We further introduce a dimensionless temperature parameter that characterizes the tendency of the flow to produce ice, the reduced temperature $\overline{T}$:

\begin{equation}
\label{eq def temperature reduite}
\overline{T}=\frac{T_m-T_s}{T_{\rm{in}}-T_m}
\end{equation}

This parameter expresses the ratio between the temperature differences of the solid and liquid phases with respect to the melting point $T_m$, thereby combining the effects of $T_\mathrm{in}$ and $T_s$ into a single nondimensional variable. A higher $\overline{T}$ corresponds to a stronger driving force for solidification.

To evaluate the influence of each control parameter on the resulting ice morphology, Figure~\ref{Macrostructure et differents etats finaux}(b–f) presents five height maps of the ice surface after $18 \mathrm h$ obtained under different experimental conditions, each highlighting the effect of varying a single control parameter relative to the reference case (a). The most prominent influence of these parameters concerns the maximum ice height, indicated by the color scale. Figures \ref{Macrostructure et differents etats finaux}(b–c) show that increasing the flow rate reduces the final height. Indeed, increasing the flow rate increases the heat flux brought by the rivulet and reduces the velocity of the solidification front. Conversely, Figures\ref{Macrostructure et differents etats finaux}(d–e) reveal that the ice height increases with $\overline{T}$, as the larger thermal contrast between the liquid and the solid enhances the solidification rate. Finally, changes in substrate inclination (Figure~\ref{Macrostructure et differents etats finaux}(f)) seem to have only a minor effect on the final height field.

All experiments exhibit a comparable two-region morphology: an upstream triangular expansion zone, followed by a region of constant width once the ice reaches the substrate edges. This largely symmetric contour reflects the extent of rivulet spreading driven by thermo-hydrodynamic coupling. The spreading can be quantified by the opening angle of the upstream triangular region. A preliminary analysis of this parameter is presented in Appendix~\ref{Appendix angle de cone}.

\subsection{Spreading of the ice structure induced by rivulet oscillations}
\label{sec spreading and rivulet oscillations}

\begin{figure}
    \centering
    \includegraphics[width=\linewidth]{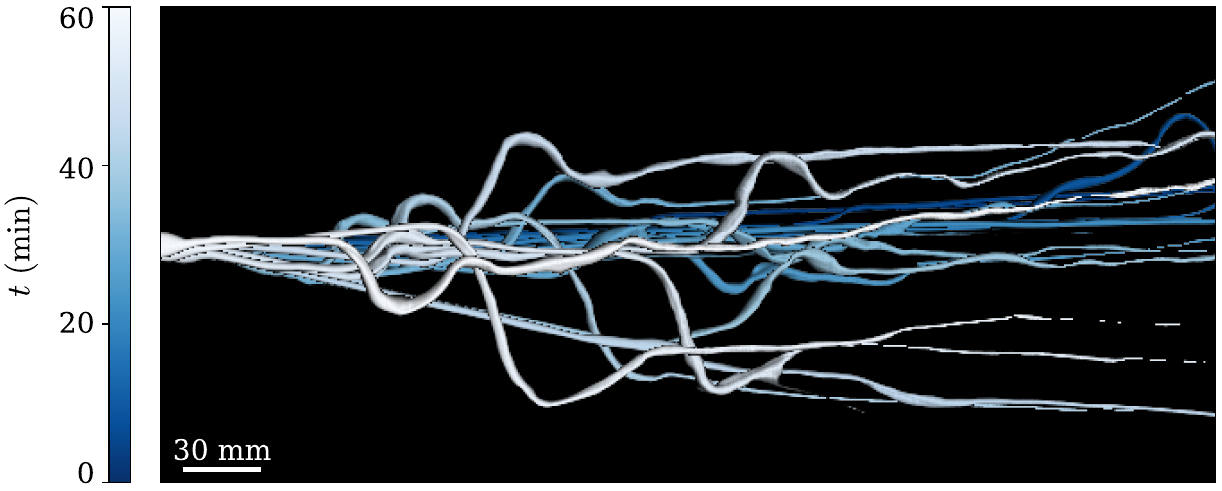}
    \caption{Evolution of the rivulet trajectory during the first hour of the reference experiment captured with the thermal camera. The colorbar indicates the detection time of the rivulet from the thermal measurements.}
    \label{Meandrage qualitatif}
\end{figure}

To better understand the evolution of the ice structure, one can analyze the motion of the rivulet over its surface using the surface temperature measurements provided by the infrared camera. Figure~\ref{Meandrage qualitatif} shows the evolution of the rivulet’s trajectory over time, at successive time steps (color-coded from blue to white), for the reference experiment. The position of the rivulet is identified through the regions with temperatures $T>0^\circ \mathrm C$.

The rivulet quickly deviates from its initial straight path and begins to oscillate irregularly as it flows over the developing ice structure. This explains the development of new ice branches highlighted on Figure~\ref{Macrostructure et differents etats finaux}(a) for $t< 1 \mathrm{h}$. Over time, the amplitude of these oscillations increases as the ice structure expands. One can understand this dynamic evolution as a change of wettability of the rivulet on the ice surface, coupled with a change in the surface topography due to ice formation. Indeed, the macroscopic contact angle of water on ice increases with the undercooling of the ice surface (ie, the difference between the solidification temperature and the ice surface temperature) due to contact line pinning \citep{Sarlin_2025, Demmenie_2025}. Thus, as the experiment progresses, the overall ice height increases and the surface temperature increases, which in turn causes the rivulet to depin and migrate laterally. As a result, the amplitude of oscillations observed after one hour of experiment is, on average, significantly greater than during the first ten minutes. This type of rivulet oscillations can be observed for all experiments.

\begin{figure}
    \centering
    \includegraphics[width=\linewidth]{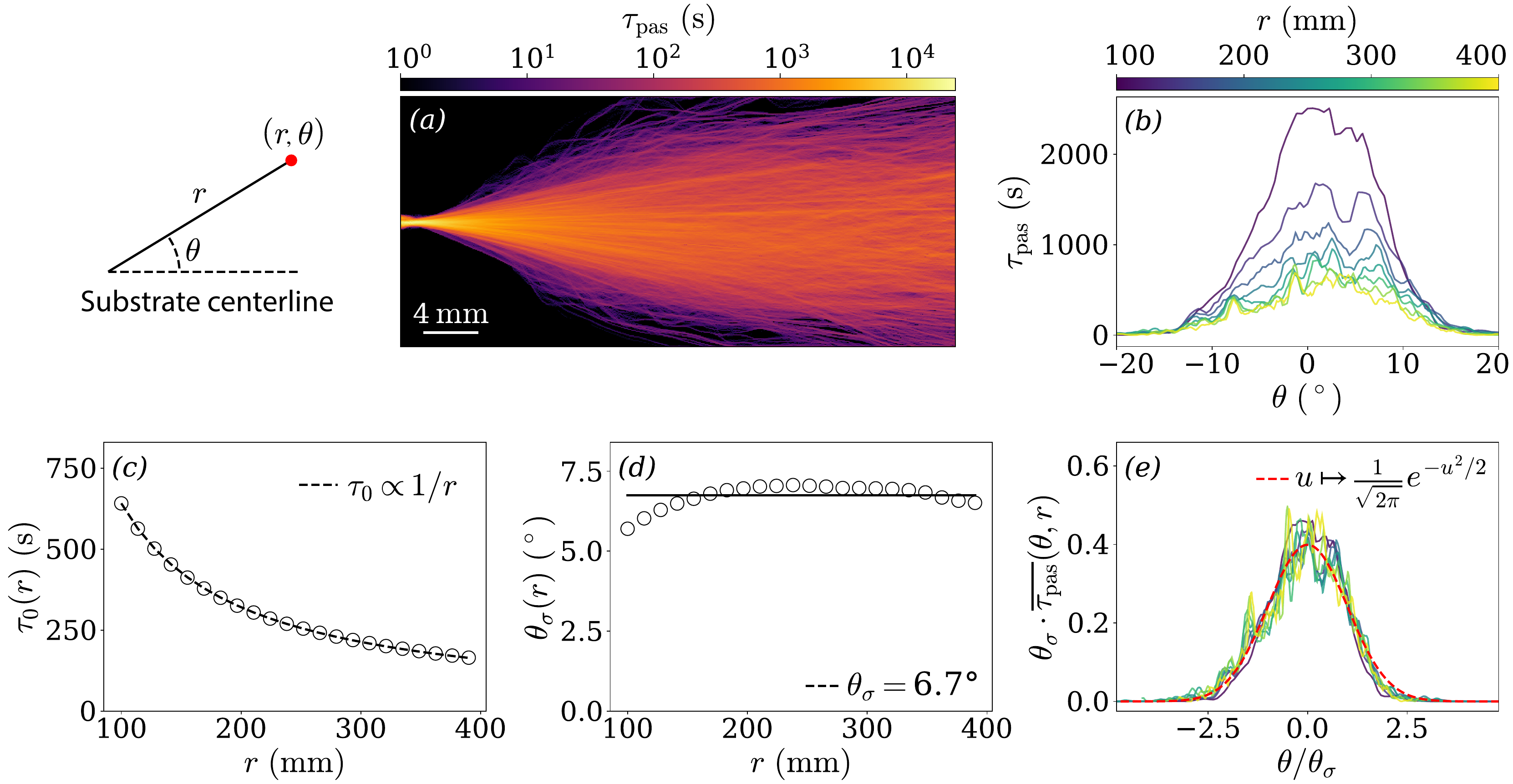}
    \caption{(a) Distribution of the total rivulet residence time $\tau_\mathrm{pas}$ over the study area for the reference experiment. Owing to the resulting geometry, the distribution is analyzed using cylindrical coordinates $(r,\theta)$ defined with respect to the substrate centerline. Panel (b) shows eight angular cross-sections of $\tau_\mathrm{pas}$ taken at different radial distances ranging from 100 to $400 \ \mathrm{mm}$. (c) Total residence time $\tau_0$ obtained by integrating $\tau_\mathrm{pas}$ over a circular arc at a given radial distance $r$. Dividing $\tau_\mathrm{pas}$ by $\tau_0$ defines the normalized residence time $\overline{\tau_\mathrm{pas}}$, which is fitted with a centered Gaussian distribution to extract the angular standard deviation $\theta_\sigma(r)$, reported in panel (d). (e) The rescaled profiles $\theta_\sigma.\overline{\tau_\mathrm{pas}}$ are plotted as a function of the reduced angle $\theta/\theta_\sigma$, where $\theta_\sigma = 6.7^\circ$ is the average standard deviation obtained from panel (d), demonstrating the self-similar nature of the residence-time distributions.}
    \label{Rivulet residence time}
\end{figure}

To further characterize the rivulet displacement over the entire duration of the experiment, we analyzed the total residence time $\tau_\mathrm{pas}$ of the rivulet at each point on the substrate. Figure~\ref{Rivulet residence time} shows the spatial distribution of $\tau_\mathrm{pas}$ for the reference experiment. Panel (a) shows the total residence time $\tau_\mathrm{pas}$ at each point of the surface. The regions shown in black correspond to points where the rivulet never passed. Owing to the geometry of the distribution, the data are analyzed in cylindrical coordinates $(r,\theta)$ defined with respect to the substrate centerline, as illustrated in the sketch at the top left of Figure~\ref{Rivulet residence time}. For a given radius $r$, $\tau_\mathrm{pas}$ is significantly larger near the center of the ice structure compared to its outer edges, highlighting a strong dependence on the azimuthal angle $\theta$. This dependence is further highlighted in panel (b), which shows $\tau_\mathrm{pas}$ as a function of $\theta$ for eight radial distances ranging from 100 to $400 \ \mathrm{mm}$. The residence time clearly reaches a maximum at $\theta = 0$ and decreases rapidly as $|\theta|$ increases. Moreover, each radial cross-section exhibits a behavior close to a Gaussian distribution, with a peak amplitude at $\theta = 0$ that decreases with increasing $r$: at larger radial distances, the distribution becomes both broader and lower. The total residence time $\tau_0$ integrated over a circular arc at a given radial distance $r$ is shown in Figure~\ref{Rivulet residence time}(c) and displays a clear decay following $\tau_0 \propto 1/r$, as emphasized by the fitted dotted line. Assuming that the radial and angular dependences of the residence time distribution can be separated, this scaling arises from the conservation of the total residence time along a circular arc, such that $\int_{\theta=-\pi/2}^{\pi/2} \tau_\mathrm{pas}(r,\theta) r \mathrm{d}\theta = \mathrm{cst}$. To quantify the angular spreading of the distribution, $\tau_\mathrm{pas}(r,\theta)$ is normalized by its angular integral $\tau_0(r)$, yielding the reduced variable $\overline{\tau_\mathrm{pas}}(r,\theta)$. This reduced distribution is then fitted by a Gaussian function characterized by a standard deviation $\theta_\sigma$:

\begin{equation}
\label{eq def gaussienne ecart type pour fit comptage}
{\cal G}(\theta,\theta_\sigma) = \frac{1}{\sqrt{2\pi}\theta_\sigma}\exp \left(-\frac{\theta^2}{2 \theta_\sigma^2}\right)
\end{equation}

As shown in Figure~\ref{Rivulet residence time}(d), the extracted angular standard deviation $\theta_\sigma$ varies only weakly with $r$ and is well approximated by a plateau with a mean value $\theta_\sigma = 6.7^\circ$. This indicates that the normalized residence time distribution follows a unique Gaussian profile characterized by a constant standard deviation $\theta_\sigma = 6.7^\circ$. To illustrate this self-similar behavior, Figure~\ref{Rivulet residence time}(e) shows the rescaled variable $\theta_\sigma \overline{\tau_\mathrm{pas}}$ as a function of the reduced azimuthal angle $\theta/\theta_\sigma$, using the mean value $\theta_\sigma = 6.7^\circ$ extracted from Figure~\ref{Rivulet residence time}(d). The collapse of the distributions obtained at different radial positions demonstrates that this Gaussian model accurately reproduces the experimental profiles. This analysis therefore shows that the rivulet residence time distribution can be described by the following self-similar expression:

\begin{equation}
\label{eq fonction theorique autosimilaire comptage}
\tau_{\rm{pas}}(r,\theta) \propto \frac{1}{r}\exp \left(-\frac{\theta^2}{2\theta_\sigma^2}\right)
\end{equation}

In conclusion, the rivulet residence time follows a centered Gaussian distribution in $\theta$, while its dependence on $r$ enters only through the prefactor associated with the arc perimeter. This indicates that the rivulet predominantly flows near the central region of the structure and only rarely reaches its lateral boundaries. This result will later prove to be key for explaining the angular dependence of the ice height field (see Section~\ref{sec angular variations}), as already suggested by the different final states highlighted in Figure~\ref{Macrostructure et differents etats finaux}. Additionally, this Gaussian distribution for the rivulet residence time is observed for all the span of experimental conditions of this study. Yet, the standard deviation $\theta_\sigma$ depends on the control parameters: the wider the lateral envelope of the ice structure the higher $\theta_\sigma$. Since the aperture of the envelope varies predominantly with the thermal parameters of the rivulet (see Appendix~\ref{Appendix angle de cone} for details), we can infer that $\theta_\sigma$ will increase with $\overline{T}$.

\subsection{Convergence of the ice structure towards a stationary state}
\label{sec convergence spatio-temporelle hauteur}

In the previous section, we analyzed the global formation of the ice structure in the reference experiment, from the initial straight ridge to the final geometry. A natural question arises: does this final configuration correspond to a stationary state? To address this, we divide the structure into two components: the lateral envelope, and the height field inside this envelope. 

\begin{figure}
    \centering
    \includegraphics[width=\linewidth]{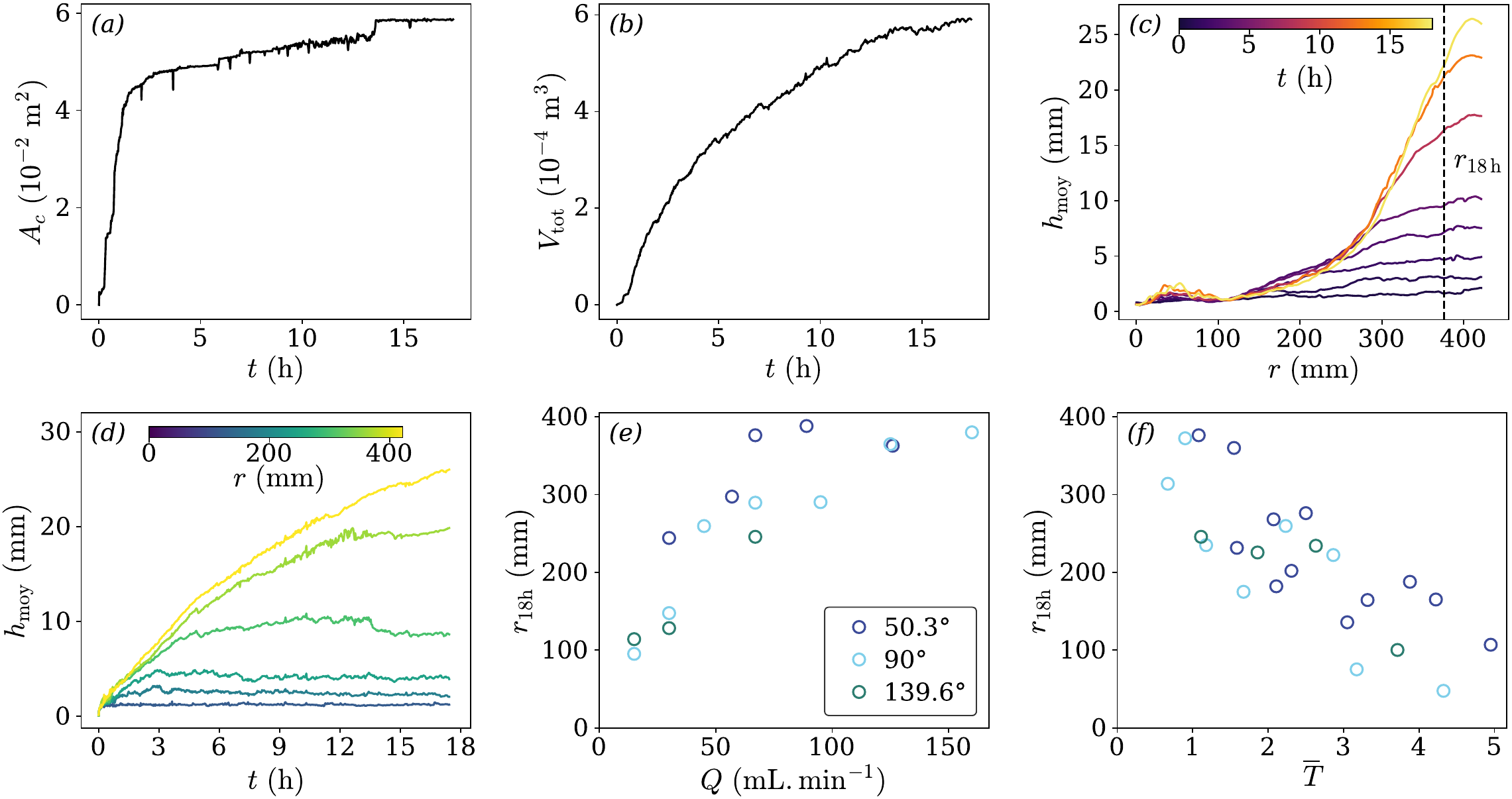}
    \caption{For the reference experiment of Figure~\ref{Macrostructure et differents etats finaux}(a), this figure shows: (a) the contact area $A_c$ of the ice structure with the substrate, (b) the total ice volume $V_\mathrm{tot}$, and (c) the angularly averaged height $h_\mathrm{moy}$ computed as a function of the radial distance $r$, at different times during the experiment. (d) The temporal evolution of $h_\mathrm{moy}$ is plotted for several values of $r$ within the observation region. From these curves we extract a convergence radius $r_\mathrm{18 h}$, defined as the largest $r$ for which $t \mapsto h_\mathrm{moy}(r,t)$ varies on average by less than $1 \ \mathrm{mm}$ during the last 4 hours of the experiment. In this example, $r_\mathrm{18 h} = 376 \ \mathrm{mm}$: the second curve from the top (light green) has reached a stationary height after $18 \mathrm{h}$, whereas the top curve (yellow) has not. (e) $r_\mathrm{18 h}$ as a function of flow rate $Q$ for constant temperature conditions ($\overline{T} = 1.1$), and (f) as a function of the reduced temperature $\overline{T}$ for a constant flow rate ($Q = 67 \ \mathrm{mL.min^{-1}}$). Each color corresponds to a different substrate inclination.}
    \label{Evolution variables principales et r18h}
\end{figure}

In order to study the convergence of the lateral envelope, we can analyze the contact area $A_c$ between the ice and the heat exchanger. This variable is illustrated on Figure~\ref{Evolution variables principales et r18h}(a) for the reference experiment. It increases rapidly during the first two hours before gradually reaching a steady value, confirming the rapid early expansion of the ice envelope observed in Figure~\ref{Macrostructure et differents etats finaux}(a). In Appendix \ref{Appendix angle de cone}, we show that the lateral envelope consistently reaches a stationary state characterized by a triangular shape, which results from a gravito-capillary balance combined with thermal effects. We further show that the opening angle depends primarily on the thermal parameters, rather than the hydrodynamic ones, indicating that phase change is the dominant mechanism shaping the external envelope of the ice structure.

Beyond the lateral envelope, we now focus on the height field within it, and in particular on whether it has reached a stationary regime by the end of the experiment. As a first glimpse at the evolution of the height field inside the lateral envelope, Figure~\ref{Evolution variables principales et r18h}(b) illustrates the temporal evolution of the total volume of ice $V_\mathrm{tot}$ formed during the reference experiment. We can observe that the rate of ice formation decreases during the experiment: the higher the ice structure the lower the heat flux absorbed from the rivulet by the ice and the slower the ice front progresses. Contrarily to $A_c$, $V_\mathrm{tot}$ has not quite reached a stationary value after 18h of experiment but keeps increasing slowly, indicating that the ice block keeps thickening during the whole period of time of the experiment and would continue to do so if the test was continued. Finally, in order to quantify the mean ice growth at different radial positions, one can define the angularly averaged height $h_\mathrm{moy}$ at time $t$ along a circular arc of radial distance $r$ as

\begin{equation}
\label{Definition hmoy}
    h_{\rm{moy}}(r,t) = \frac{\int_{\theta=-\pi/2}^{\pi/2} h(r,\theta,t)\mathds{1}_\mathrm{ice}(r,\theta,t)\ d\theta}{\int_{\theta=-\pi/2}^{\pi/2} \mathds{1}_\mathrm{ice}(r,\theta,t)\ d\theta}
\end{equation}

where $\mathds{1}_\mathrm{ice}(r,\theta,t)$ denotes the indicator function equal to 1 when the point lies on the ice surface and 0 when it lies on the thermal exchanger. The evolution of $h_\mathrm{moy}$ is shown in Figure~\ref{Evolution variables principales et r18h}(c). Each curve corresponds to a distinct time during the experiment, ranging from $27 \ \mathrm{min}$ to $18 \mathrm h$, as indicated by the colorbar on the top of the graph. The height is observed to increase both with time and with radial distance, as the water becomes colder farther from the injector, thus facilitating ice formation. The height field eventually reaches a steady value for $r \in [0, 376]\ \mathrm{mm}$, indicating that ice growth ceases in this region. For $r > 376\ \mathrm{mm}$, however, the ice height has not yet converged to a stationary value, showing that the ice structure would continue to grow there even after $18 \mathrm h$ of experimentation. Thus a convergence radius $r_\mathrm{18 h}$ appears to exist, such that after $18 \mathrm{h}$, the angular-averaged height field $h_\mathrm{moy}$ has converged only in the region $r < r_\mathrm{18 h}$.

To determine the convergence radius $r_\mathrm{18 h}$ quantitatively, we examine the temporal evolution of $h_\mathrm{moy}$ at several radial positions $r$, as shown in Figure~\ref{Evolution variables principales et r18h}(d) for the reference experiment. The results reveal a clear dependence on $r$: at small radial distances, $h_\mathrm{moy}$ tends toward a steady value, whereas at the largest distance, it continues to increase at $t = 18\mathrm{h}$, indicating that growth would persist beyond the experiment duration. To distinguish converged from non-converged regions, we define the following convergence criterion: for a given $r$, $h_\mathrm{moy}(r,t)$ is considered converged if the mean variation over the last $4\mathrm{h}$ of the experiment is less than $1\ \mathrm{mm}$. In this example, the outermost curve at $r = 421\ \mathrm{mm}$ does not meet this threshold, with a height increase of about $2.50\ \mathrm{mm}$ during the final 4 hours. In contrast, the curve at $r = 361\ \mathrm{mm}$ increases by only $0.58\ \mathrm{mm}$ over the same period and therefore satisfies the convergence requirement. This indicates that $r_\mathrm{18 h}$ lies between these two radial positions. We then define $r_\mathrm{18 h}$ as the largest radial distance for which the convergence criterion is verified. Applying this procedure yields $r_\mathrm{18 h} = 376\ \mathrm{mm}$ in this example.

Using this procedure, we can determine the value of $r_\mathrm{18 h}$ for each experiment and plot the results as functions of the flow rate $Q$ and the reduced temperature $\overline{T}$, as shown in Figures\ref{Evolution variables principales et r18h}(e-f). Some experiments have converged on the whole area of study (which implies that $r_\mathrm{18 h}>L_\mathrm{study}$, $L_\mathrm{study}$ being the substrate length) and are thus not presented on both graphs. We can infer from these two graphs that $r_\mathrm{18 h}$ increases with the flow rate and decreases with the reduced temperature. Furthermore, as was shown in Figure~\ref{Macrostructure et differents etats finaux}, the maximum ice height of the structure presents the opposite trends: it decreases with $Q$ and increases with $\overline{T}$. Hence we can deduce that the convergence radius decreases with the overall height field of the structure. Having established these trends for the convergence radius, we now turn to the height field within the converged region $r<r_\mathrm{18 h}$ in order to analyze the evolution of the stationary height profile.

\section{Theoretical model for the stationary height field of the ice structure}
\label{sec theoretical model}

\begin{figure}
    \centering
    \includegraphics[width=\linewidth]{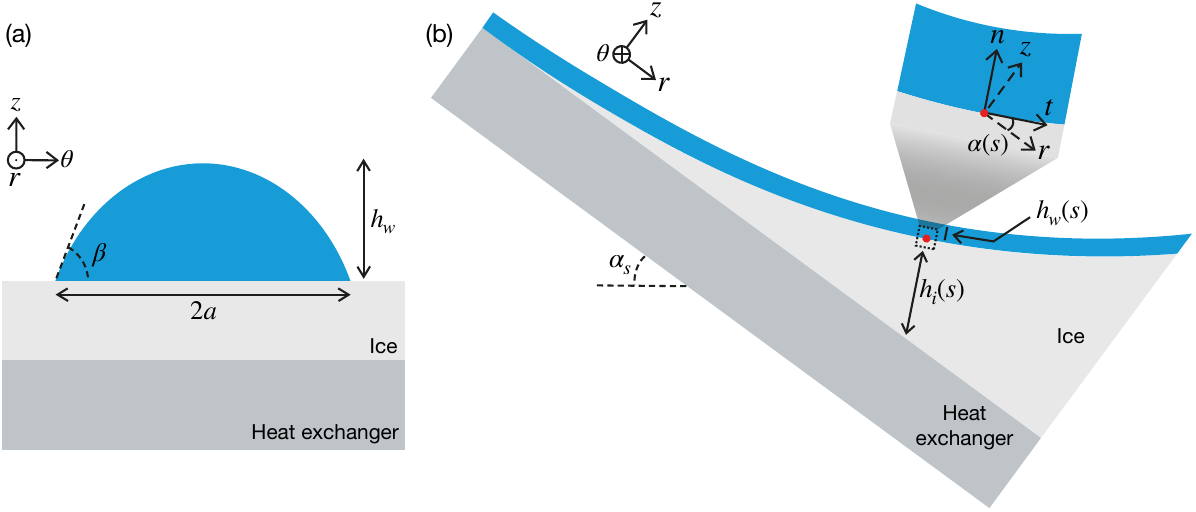}
    \caption{(a) Transverse rivulet geometry according to \citet{Duffy_1995}, with $h_w$ the maximum rivulet height, $a$ its half-width and $\beta$ its contact angle. (b) Bidimensional geometry of the rivulet flowing over the stationary ice structure. The reference frame is changed from cylindrical coordinates $(r,z)$ to curvilinear coordinates $(s,n)$ attached to the ice–water interface. The local slope of this interface is given by $\alpha_s - \alpha(s)$, with $\alpha(s) \ll 1$ related to the ice height field $h_i$ through $\tan \alpha = \nabla h_i$.}
    \label{Geometrie du modele}
\end{figure}

To describe the stationary geometry of the height field associated with our three-dimensional ice structures, we propose a generalization of the model developed by \citet{Huerre_2021}, who studied the solidification of a rivulet on a two-dimensional ice ridge at lower flow rates and smaller length scales. In that study, the authors assumed that the ice has a constant slope and that the rivulet velocity field is directed along a fixed direction, thereby neglecting any advection in the direction perpendicular to the slope. In the present work, we adapt this framework to the case of a straight rivulet flowing over a three-dimensional ice structure with a variable slope, extending its validity to larger flow rate ranges and longer spatial and temporal scales. The system is schematized on Figure~\ref{Geometrie du modele}(a). It is described using a two-dimensional model in which the $(r,z)$ plane represents a slice of the ice structure parallel to the flow direction, with an arbitrary angle $\theta$ between the flow and the centerline of the heat exchanger.

\subsection{The rivulet geometry}

Before considering a substrate of varying slope, one should provide a hydrodynamic description of the rivulet running down a substrate of constant inclination $\alpha_s$. Among the numerous models available so far to describe this rivulet geometry \citep{Towell_1966, Bentwich_1976, Duffy_1995, Perazzo_2004}, we used in this study the approach of \citet{Duffy_1995} due to its formulation that is applicable for inclinations $\alpha_s \in [0,180^\circ]$ and the relevance of the lubrication approximation for the flow (given the low contact angle of the rivulet $\beta \lesssim 40^\circ$). In their approach, the fluid is considered Newtonian and incompressible, with density $\rho$, viscosity $\mu$, and surface tension $\gamma$. The flow satisfies the thin-film conditions, and can be described by a semi-Poiseuille velocity profile and the transverse height profile is found to be a symmetric hyperbolic function, linking the local height to the rivulet contact angle $\beta$, the reduced capillary length $l_c=\sqrt{\gamma/\rho g \cos\alpha_s}$, and the Bond number $\mathrm{Bo}=a/l_c$, with $a$ the rivulet half-width (see Figure~\ref{Geometrie du modele}(a)). According to their analysis, the averaged height of the rivulet along the transverse axis obeys the following equation:

\begin{equation}
\label{eq hw DM}
\frac{h_w}{l_c \tan \beta} =
\begin{cases}
\displaystyle \coth \mathrm{Bo} - 1/\mathrm{Bo} & \quad \alpha_s < 90^\circ \\[0.3em]
\displaystyle \mathrm{Bo}/3 & \quad \alpha_s = 90^\circ \\[0.3em]
\displaystyle 1/\mathrm{Bo} - \cot \mathrm{Bo} & \quad \alpha_s > 90^\circ
\end{cases}
\end{equation}

In the case of small Bond numbers, it can be shown that this equation can be coupled with the mass conservation equation in order to obtain the following scaling for the mean height of the rivulet:

\begin{equation}
\label{eq scaling ruisselet}
h_w \propto \left(\frac{Q}{\sin \alpha_s} \right)^{1/4}
\end{equation}

In the present experiments, we can determine its average width and infer the corresponding Bond numbers using the thermal camera to detect the rivulet. This analysis shows that $\mathrm{Bo}$ spans the range $[0.4, \, 0.9]$ across all experimental conditions. We can therefore reasonably conclude that $\mathrm{Bo} \lesssim 1$, and that Equation~\eqref{eq scaling ruisselet} is applicable to our configuration of a rivulet flowing over an ice surface.

We then extend this constant-slope rivulet model to an ice structure whose slope varies as $\alpha_s - \alpha(r)$, where $\alpha = \nabla h_i$ represents the local change in slope induced by the ice creation on the substrate. Because the ice structure remains weakly inclined (the typical slope being $\nabla h_i \sim \frac{35 \ \mathrm{mm}}{420 \ \mathrm{mm}} \sim 8 \times 10^{-2} \ll 1$) we consider that Equations~\eqref{eq hw DM} and \eqref{eq scaling ruisselet} remain valid and can be applied by replacing the constant slope $\alpha_s$ with the locally varying slope $\alpha_s - \alpha(r)$. This approximation is supported by the large separation of scales between the characteristic length of ice height variation, on the order of $4 \ \mathrm{m}$, and the characteristic distance required for the establishment of the semi-Poiseuille velocity regime, on the order of $2 \ \mathrm{cm}$.

\subsection{Advected heat equation and boundary conditions on a slowly varying slope}

Then, we change the referential of study towards the curvilinear referential associated with the ice slope. Hence, we adapt the initial $(r,z)$ variables to $(s,n)$ variables, with $s(r)~=~\int_0^{r}\sqrt{1 + \nabla h_i(r')^2} dr'$ and $n$ the normal distance from the point of study to the ice/water interface. The curvilinear geometry is detailed on Figure~\ref{Geometrie du modele}(a). In this new referential, the rivulet height is a slowly varying function $h_w(s)$ and its velocity $U(s,n)$ exhibits a semi-Poiseuille profile:

\begin{equation}
\label{eq semi poiseuille curviligne}
U(s,n) = \dfrac{g \sin(\alpha_s-\alpha(s)) h_w(s)^2 }{2 \nu} \dfrac{n}{h_w(s)}\left(2-\dfrac{n}{h_w(s)}\right)
\end{equation}

with $\nu$ the kinematic viscosity of the flow. With this curvilinear referential along with this velocity profile, we can adapt the hypothesis and equations of the model by \citet{Huerre_2021}. The stationary advection-diffusion heat equation for the water temperature $T_w$ as well as the stationary diffusion equation for the ice temperature $T_i$ can be written as:

\begin{equation}
\label{eq chaleur deux phases}
\begin{cases}
    U(s,n) \dfrac{\partial T_w}{\partial s} = D_w\nabla^2 T_w \\[0.4em]
    \nabla^2 T_i = 0
\end{cases}
\end{equation}

with $D_w$ the water heat diffusion coefficient. The two variables $T_w$ and $T_i$ are linked at the ice-water interface by the stationary Stefan condition:

\begin{equation}
\label{eq condition de Stefan}
    (k_i\nabla T_i-k_w\nabla T_w).\mathbf{e}_n(s,0)=0
\end{equation}

with $k_i$ and $k_w$ the thermal conductivities of ice and water, respectively, and $\mathbf{e}_n$ the unit vector normal to the ice–water interface. Additionally, at the free surface of the rivulet ($n=h_w(s)$), we can neglect the heat influx of ambient air due to its little thermal conductivity in comparison with water ($k_\mathrm{air}/k_w \sim 0.04 \ll 1$), thus providing the equation:

\begin{equation}
\label{eq flux nul avec air}
    k_w\nabla T_w.\mathbf{e}_n(s,h_w(s))=0
\end{equation}

\begin{figure}
    \centering
    \includegraphics[width=\linewidth]{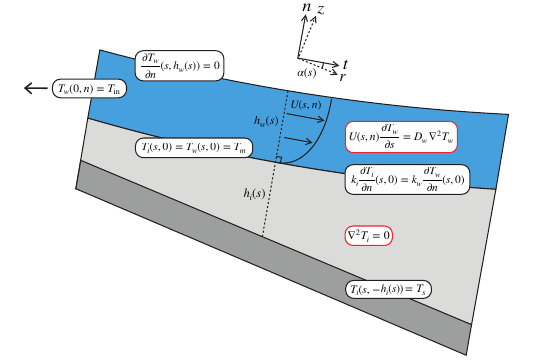}
    \caption{Schematic representation of the transverse section of a rivulet flowing over an ice structure with spatially varying slope; the heat equations in the liquid and solid are shown in red, and the boundary conditions are shown in black.}
    \label{Hypotheses modele}
\end{figure}

Additionally, both temperatures coincide with the solidification temperature of water $T_m$ at the interface, such that $T_i(s,0)=T_w(s,0)=T_m$. Finally, we assume that the injected water is at a constant temperature $T_\mathrm{in}$ and that the heat-exchanger surface remains at a constant temperature $T_s$. All model assumptions are summarized in Figure~\ref{Hypotheses modele}.

\subsection{Model solution}

Following this set of equations and boundary conditions, we can define the initial Péclet number $\mathrm{Pe}$ using the initial height of the rivulet $h_{w,0}$ (corresponding to the mean height of a rivulet flowing on top of an incline of constant slope $\alpha_s$ according to \citep{Duffy_1995}) and the initial velocity of the free surface $U_0 = g h_{w,0}^2\sin \alpha_s  /2 \nu$:

\begin{equation}
    \mathrm{Pe} = \frac{U_0 h_{w,0}}{D_w} = \frac{g h_{w,0}^3\sin \alpha_s}{2 \nu D_w}
\end{equation}

This Péclet number is very large in our experiments, spanning roughly from 600 to 4000, indicating that heat transport is governed predominantly by advection rather than diffusion. Then we can use the dimensionless set of parameters:

\begin{equation}
\begin{cases}
    \overline{s} = \dfrac{s}{h_{w,0}\, \mathrm{Pe}}, \quad \overline{n} = \dfrac{n}{h_w}, \quad \overline{h_i}= \dfrac{h_i}{h_{w,0}}, \quad \overline{h_w}= \dfrac{h_w}{h_{w,0}} \\[1em]
    \Theta_w = \dfrac{T_w-T_m}{T_\mathrm{in}-T_m}, \quad \Theta_i = \dfrac{T_i-T_m}{T_s-T_m}
\end{cases}
\end{equation}

From the previous set of equations and the newly defined dimensionless variables, and considering $\mathrm{Pe} \gg 1$, it can be shown that the dimensionless temperature fields take the following form after neglecting terms of order 1 or more in $1/\mathrm{Pe}$ (see Appendix \ref{Appendix detail obtention equation theorique hauteur glace} for the detailed derivation):

\begin{equation}
    \begin{cases}
        \Theta_i(\overline{s},\overline{n}) = 1+\overline{n}\dfrac{\overline{h_w}(\overline{s})}{\overline{h_\mathrm{i}}(\overline{s})} \\[1em]
        \Theta_w(\overline{s},\overline{n}) = A_1\Phi_1(\overline{n})e^{-\lambda_1^2 \overline{s}}
    \end{cases}
\end{equation}

where $A_1 \approx 0.78$, $\lambda_1 \approx 1.68$, and $\Phi_1$ denotes the first eigenfunction associated with the corresponding Sturm–Liouville problem, whose first eigenvalue is $\lambda_1$. The details of the Sturm-Liouville decomposition can be found in \cite{Huerre_2021}. Then, by applying the Stefan condition (Equation~\eqref{eq condition de Stefan}) together with the relationship between the curvilinear coordinate $s$ and the radial distance $r$, and retaining terms down to order 0 in $\mathrm{Pe}$, the dimensional stationary height of the ice structure can be expressed as follows:

\begin{equation}
\label{eq hauteur glace theorique}
\begin{cases}
        h_i(r) =  h_0\exp(\dfrac{r}{r_c}) \\[0.8em]
        h_0 = \dfrac{k_i \overline{T}h_{w,0}}{k_w A_1 \Phi_1'(0)}, \quad r_c = \dfrac{h_{w,0} \rm{Pe}}{\lambda_1^2}
\end{cases}
\end{equation}

Although an exponential growth was already predicted by \citet{Huerre_2021}, the present approach incorporates a hydrodynamic model of the rivulet, allowing for a more precise link between the Péclet number and the flow rate, and further extends their results to ice surfaces with varying slopes, as encountered in our experiments. This theoretical exponential profile is consistent with the experimental evolution of the angularly averaged height $h_\mathrm{moy}$ at the end of the experiment, as illustrated in Figure~\ref{Evolution variables principales et r18h}(c) for $r < r_\mathrm{18 h}$ at $t = 18 \mathrm{h}$. Using the hydrodynamic relations for $h_{w,0}$ and $\mathrm{Pe}$ derived from Equation~\eqref{eq scaling ruisselet}, we find that the theoretical ice height parameters $h_0$ and $r_c$ can be expressed as:

\begin{equation}
\label{eq hauteur theorique avec parametres de controle}
h_0 = 0.44 \frac{k_i}{k_w} \left(\frac{\nu \tan \beta}{g}\right)^{1/4} \overline{T} \left(\frac{Q}{\sin \alpha_s}\right)^{1/4}
        , \quad r_c = 0.057 \frac{Q \tan \beta}{D_w}
\end{equation}

Hence, among the control parameters varied in our experiments, the characteristic length of the exponential profile, $r_c$, depends predominantly on the hydrodynamic properties of the flow, while the initial height is mainly governed by the thermal conditions through $\overline{T}$. Interestingly, the substrate inclination has only a weak influence on the initial ice height $h_0$, and does not affect the characteristic distance $r_c$. The rivulet–ice contact angle has a small effect on the prefactor associated with $h_0$, but plays a more significant role in the prefactor of $r_c$.

\section{Results}

We have shown in the previous sections that the ice structure converges in a delimited region $r<r_\mathrm{18 h}$ and we have constructed a two dimensional theoretical model based on a three dimensional rivulet geometry and the resolution of the heat equations in both the liquid and solid domains, leading to Equations~\eqref{eq hauteur glace theorique} and \eqref{eq hauteur theorique avec parametres de controle}. We can now compare these theoretical predictions with the experimental height field.

\subsection{Ice structure height on the heat exchanger centerline}

To test the validity of Equation~\eqref{eq hauteur glace theorique} for the stationary ice height, we first examine the centerline of the heat exchanger — the region most frequently visited by the rivulet during its lateral oscillations, as revealed by the residence time analysis in Figure~\ref{Rivulet residence time}. Based on the convergence analysis presented in Section~\ref{sec convergence spatio-temporelle hauteur} and the definition of the convergence radius $r_\mathrm{18 h}$, we compare the experimental height profile along the central axis at $t = 18  \mathrm{h}$ with the theoretical expression given in Equation~\eqref{eq hauteur glace theorique}. To this end, we fit the experimental profile with an exponential function, from which two extrapolated parameters, $h_0$ and $r_c$, are obtained.

\begin{figure}
    \centering
    \includegraphics[width=\linewidth]{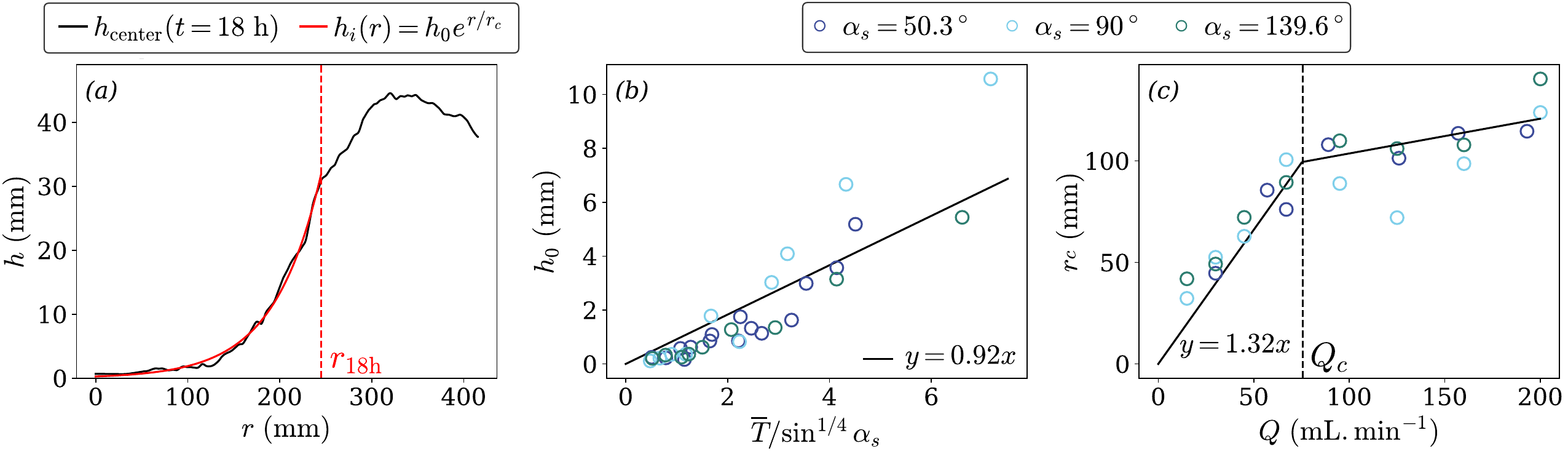}
    \caption{(a) For $\alpha_s = 50.3^\circ$, $Q = 30 \ \mathrm{mL.min^{-1}}$, and $\overline{T} = 1.07$, comparison between the experimental ice height measured along the centerline of the substrate at $t = 18  \mathrm{h}$ and its fitted exponential profile $h_i$ (as suggested by Equation~\eqref{eq hauteur glace theorique}) in the convergence region $r < r_\mathrm{18 h} = 245 \ \mathrm{mm}$. This fit allows extracting both a characteristic distance $r_c$ and an initial height $h_0$ from the experiment. (b) Evolution of the extrapolated parameter $h_0$ with the reduced parameter $\overline{T}/\sin^{1/4}\alpha_s$ for a constant flow rate $Q = 67 \, \mathrm{mL.min^{-1}}$. The black line represents a linear fit to the data points. (c) Evolution of the extrapolated parameter $r_c$ with the flow rate $Q$ for a constant reduced temperature $\overline{T}=1.1$. The black line again represents a linear fit to the data points up to a critical flow rate $Q_c$, beyond which the slope changes for $Q > Q_c$, highlighting the transition in hydrodynamic regime from a straight rivulet to meanders.}
    \label{Comparison height with model}
\end{figure}

An example of the exponential fit is shown in Figure~\ref{Comparison height with model}(a), which displays an excellent agreement with the experimental data in the convergence region $r < r_\mathrm{18 h} = 245 \ \mathrm{mm}$, validating the relevance of the theoretical model. Using this exponential fitting procedure, we extract the experimental parameters $h_0$ and $r_c$. The evolution of $h_0$ with the reduced parameter $\overline{T}/\sin^{1/4}\alpha_s$ is shown in Figure~\ref{Comparison height with model}(b), while the evolution of $r_c$ with the flow rate $Q$ is presented in Figure~\ref{Comparison height with model}(c). Both plots include the three tested substrate inclinations $\alpha_s \in \{ 50.3^\circ, \, 90^\circ, \, 139.6^\circ \}$. We first observe that $h_0$ increases linearly with $\overline{T}/\sin^{1/4}\alpha_s$, with low dispersion among the different inclinations, confirming the theoretical prediction of Equation~\eqref{eq hauteur theorique avec parametres de controle}. A linear fit yields an experimental slope of approximately $0.92 \ \mathrm{mm^{-1}}$. To compare this slope with the theoretical prediction of Equation~\eqref{eq hauteur theorique avec parametres de controle}, a value of the rivulet contact angle (as defined in Equation~\eqref{eq hw DM}) must be specified. According to the model of \citet{Duffy_1995}, the rivulet half-width $a$ is related to its contact angle $\beta$ through $a \propto \tan^{-3/4}\beta$, which holds for small Bond numbers. Therefore, an estimate of $\beta$ can be obtained from measurements of the rivulet width. By analyzing the average width over the full duration of three experiments performed at different flow rates and $T_s = -22^\circ \mathrm C$, we obtain $\beta \approx 32^\circ$, which is consistent with the value reported by \citet{Sarlin_2025} for an undercooling of about $22\ \mathrm{K}$. Using this estimate for the contact angle, Equation~\eqref{eq hauteur theorique avec parametres de controle} predicts a theoretical slope of $4.39~\mathrm{mm^{-1}}$, i.e. about 4.8 times larger than the experimental prefactor. This discrepancy likely arises from several sources: the transverse averaging intrinsic to the two-dimensional model, the truncation of the theoretical solution to the first mode of the Sturm–Liouville decomposition \citep{Huerre_2021}, and the sensitivity of the exponential fitting procedure, which leads to larger relative uncertainties in the prefactor compared to the characteristic distance $r_c$.

On the other hand, Figure~\ref{Comparison height with model}(c) illustrates the experimental evolution of the characteristic length $r_c$, as defined by Equation~\eqref{eq hauteur glace theorique}. The graph exhibits two distinct regimes: a first region for $Q \in [15, 70] \ \mathrm{mL.min^{-1}}$, where $r_c$ increases linearly with $Q$, and a second region for $Q \in [70, 200] \ \mathrm{mL.min^{-1}}$, where $r_c$ continues to increase linearly but much more slowly. These two regimes are separated by a clear change in slope around $Q \sim 70 \ \mathrm{mL.min^{-1}}$. Moreover, at a given flow rate, the three substrate inclinations show once again very little dispersion. These observations are consistent with the theoretical expression of $r_c$ given in Equation~\eqref{eq hauteur theorique avec parametres de controle}, which predicts a linear dependence on $Q$ at low flow rates and no dependence on $\alpha_s$. The change in slope and the weak increase of $r_c$ for $Q > 70 \ \mathrm{mL.min^{-1}}$ can be attributed to a transition in the hydrodynamic regime: the rivulet is no longer “straight,” as assumed in the model of \citet{Duffy_1995}, but begins to meander and move laterally. This transition likely alters the assumptions of the model — particularly the rivulet’s cross-sectional geometry and contact angle, which directly influence Equation~\eqref{eq scaling ruisselet} — and consequently affects the final expression of Equation~\eqref{eq hauteur theorique avec parametres de controle}. Additionally, the experimental prefactor in the first region ($r_c=1.32.Q$) is 1.8 times higher than the theoretical prefactor, which again confirms the relevance of the theoretical model. 

\subsection{Angular variations of the ice height field}
\label{sec angular variations}

In the previous analysis of the ice height field, we demonstrated the convergence of the central height profile $h_\mathrm{center}$ toward a stationary shape consistent with the theoretical model described in Section~\ref{sec theoretical model}. However, it remains to be determined whether this convergence also holds at nonzero angular positions $\theta$, away from the centerline. In the ideal limit of infinitely long experiments, one would expect an axisymmetric ice structure, invariant with respect to $\theta$ and increasing exponentially with the radial distance $r$, since the rivulet residence time would then be uniformly infinite within the lateral envelope of the structure. This would ensure convergence of the ice height inside this envelope toward the $\theta$-independent theoretical profile derived in Section~\ref{sec theoretical model}. In practice, however, the finite duration of the experiments leads to a pronounced angular dependence of the height field, as already suggested by the different final states shown in Figure~\ref{Macrostructure et differents etats finaux}.

To interpret the angular variations observed in the ice structures, it is essential to recall the angular distribution of the rivulet residence time over the heat exchanger. As shown in Figure~\ref{Rivulet residence time}, the residence time follows a Gaussian profile in $\theta$, while its dependence on $r$ enters only through its prefactor. This implies that, even if the solidification rate and the rivulet velocity field were strictly independent of $\theta$, the fact that the rivulet spends much less time near the lateral edges than along the centerline will inevitably influence the final geometry of the ice after 18h of growth. Consequently, the cross-section of the ice at a fixed radial distance results from the convolution of two effects: the Gaussian distribution of the rivulet residence time in $\theta$, and the transient evolution required for the height field to reach its stationary regime, as described by the theoretical model. If the convergence time at the centerline is long compared to the overall duration of the experiment, the structure remains in a transient regime over most of the angular extent, and the height distribution retains a strong Gaussian signature. Conversely, if the convergence time is short, a wider central region reaches its stationary height and the cross-section becomes increasingly independent of $\theta$.

\begin{figure}
    \centering
    \includegraphics[width=\linewidth]{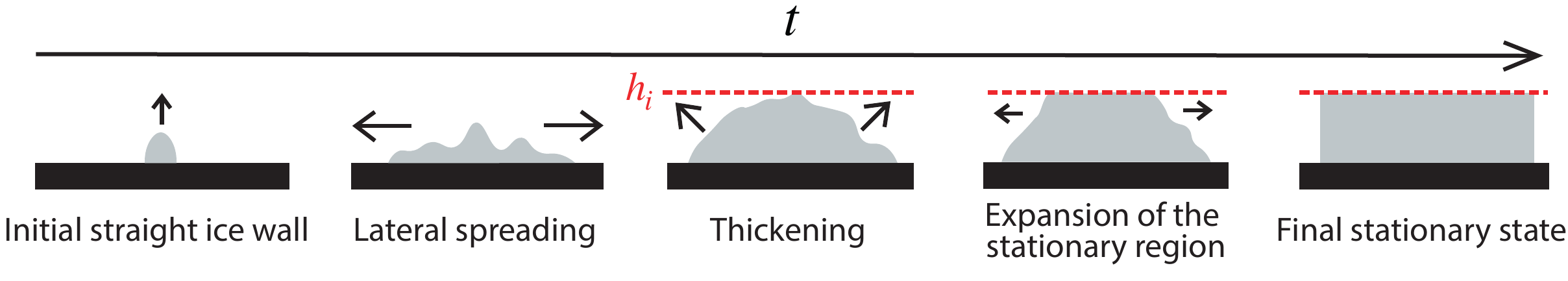}
    \caption{Schematic illustrating the evolution of an angular cross-section located at a distance $r$ from the injection point. The initially narrow and rectilinear ridge first spreads laterally before the structure thickens and converges in height toward a stationary value $h_i$ along the central line. The angular width of the converged region then gradually increases until a final stationary, rectangular cross-section is reached, for which the height field becomes invariant with $\theta$.}
    \label{Evolution of an ice cross section}
\end{figure}

From this reasoning, we can infer the overall evolution of an ice cross section at a fixed radial distance $r$, as schematized in Figure~\ref{Evolution of an ice cross section}. The experiment begins with the formation of a narrow, rectilinear ice ridge along the central axis. After some time, the rivulet flowing on top of the ridge destabilizes and deviates from the centerline, spreading laterally over the cooled substrate and thus broadening the ice structure. Subsequently, the ice thickens until it reaches the stationary height $h_i(r)$, first attained at $\theta = 0$, where the rivulet residence time is maximal. Depending on the experimental conditions and the radial position, this stationary regime may not be attained within the $18 \mathrm h$ duration of the experiment. Once this theoretical height is reached locally, the stationary region where $h = h_i(r)$ progressively extends laterally, forming a plateau centered on the heat exchanger axis. In the limit of an infinitely long experiment, this lateral expansion would ultimately lead the cross section to converge toward a rectangular profile.

\begin{figure}
    \centering
    \includegraphics[width=\linewidth]{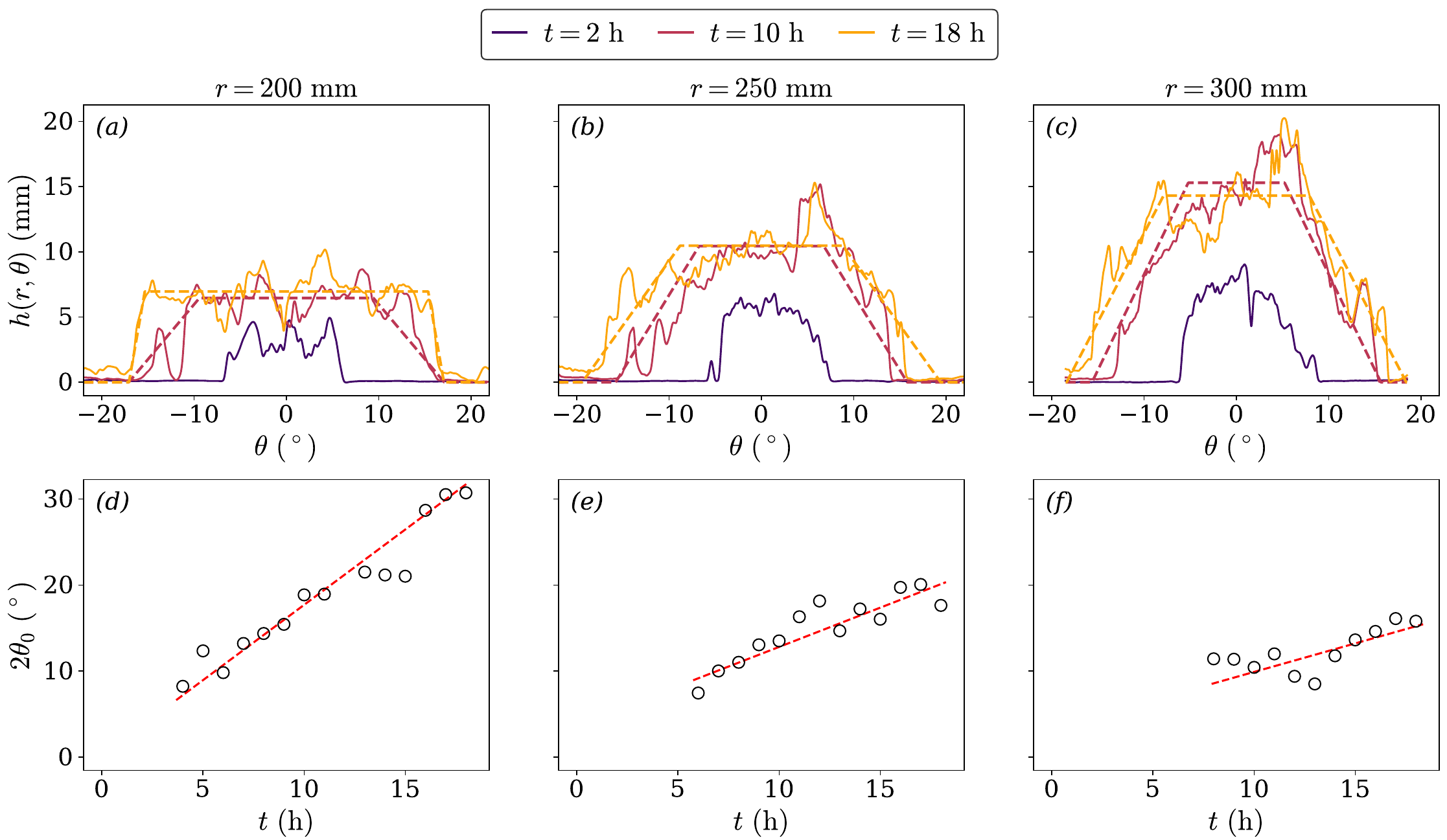}
    \caption{Lateral expansion of the converged height for an experiment under the conditions $\alpha_s=50.3^\circ$, $Q=200 \ \mathrm{mL.min^{-1}}$ and $\overline{T}=3.17$. (a–c) Three angular cross-sections recorded at different times of the experiment $t \in \{2, \, 10, \, 18\} \mathrm{h}$, each corresponding to a distinct radial distance $r \in \{200, \, 250, \, 300\} \ \mathrm{mm}$. The dashed lines correspond to piecewise linear fits of the profiles, with a central plateau corresponding to the converged region. (d–f) Temporal evolution of the plateau width $2\theta_0$ corresponding to a converged height. From these plots, a linear increase in the plateau’s angular width is observed, allowing us to estimate its growth rate. The red dashed lines in panels (d–f) indicate linear trends, with respective slopes of $1.6^{\circ}\!/\mathrm{h}$, $0.9^{\circ}\!/\mathrm{h}$ and $0.7^{\circ}\!/\mathrm{h}$.}
    \label{Exemple epatement}
\end{figure}

To quantify this final stage of lateral spreading within the converged region, we examined angular profiles of the ice height at fixed radial distances. Figure~\ref{Exemple epatement} presents a representative example for an experiment performed at $\alpha_s = 50.3^\circ$, $Q = 200,\mathrm{mL,min^{-1}}$, and $\overline{T} = 3.17$. Panels (a–c) display the angular variation of the ice height field at three different times $t \in \{2, \, 10, \, 18\} \mathrm{h}$ and for three radial distances $r \in \{200, \, 250, \, 300\} \ \mathrm{mm}$. A clear thickening and broadening of the cross-section with time is observed. The broadening of the converged region around the center of each cross-section occurs once the central height has reached its stationary value, which occurs after approximately 3h for $r = 200 \ \mathrm{mm}$, 5h for $r = 250\ \mathrm{mm}$, and 8h for $r = 300\ \mathrm{mm}$. In addition, because the stationary height increases exponentially with $r$, the profiles at $r = 200\ \mathrm{mm}$ appear much more $\theta$-independent than those at $r = 250\ \mathrm{mm}$ and $r = 300\ \mathrm{mm}$. At smaller radial distances, the stationary height is reached more rapidly, so the convolution with the Gaussian distribution of the rivulet residence time yields a wider angular region that has converged at the end of the experiment. Conversely, at larger $r$, the slower convergence maintains a stronger $\theta$-dependence in the cross-section at $t=18 \mathrm h$.

% To further our analysis of this spreading, once the ice height at the centerline has converged towards a stationary value (ie after 3h, 5h and 8h for $r\in\{200, \, 250, \, 300\} \ \mathrm{mm}$, respectively), each cross-section was fitted with a simple “plateau” function: a constant height for $|\theta| < 2\theta_0$, where $2\theta_0$ denotes the angular width of the plateau, linearly connected to zero height away from the centerline. Fitting the height profile with this "plateau" function allows us to estimate the angular width $2\theta_0$ over which the ice height is approximately constant and has reached a stationary state. These fits, reported as dashed lines in Figures \ref{Exemple epatement}(a–c), reveal that the plateau width increases steadily over time for all examined radii. This fitting procedure was repeated every hour after the height field has converged at its centerline, for $r = 200$, $250$ and $300 \ \mathrm{mm}$. The temporal evolution of this angle is shown in Figures \ref{Exemple epatement}(d–f). In each case, $2\theta_0(t)$ increases roughly linearly with time, and its growth rate decreases with increasing $r$. From linear fits, the angular spreading rates are estimated as $1.6^{\circ}\!/\mathrm{h}$ for $r = 200 \ \mathrm{mm}$, $0.9^{\circ}\!/\mathrm{h}$ for $r = 250 \ \mathrm{mm}$, and $0.7^{\circ}\!/\mathrm{h}$ for $r = 300 \ \mathrm{mm}$ (red dashed lines). 

To further analyze this spreading, once the ice height at the centerline has converged to a stationary value (i.e. after 3, 5, and 8h for $r \in \{200, \, 250,\, 300 \} \ \mathrm{mm}$, respectively), each cross-section is fitted with a simple “plateau” function. This function consists of a constant height for $|\theta| < \theta_0$ (so that $2\theta_0$ defines the angular width of the plateau), linearly connected to zero height away from the centerline. Fitting the height profiles with this plateau function allows us to estimate the angular extent $2\theta_0$ over which the ice height is approximately constant and has reached a stationary state. These fits, shown as dashed lines in Figures~\ref{Exemple epatement}(a–c) for the plots corresponding to $t=10 \mathrm h$ and $t=18 \mathrm h$, indicate that the plateau width increases steadily with time for all examined radii. This fitting procedure is repeated every hour after convergence of the central height, for $r = 200$, $250$ and $300 \ \mathrm{mm}$. The temporal evolution of $2\theta_0$ is reported in Figures~\ref{Exemple epatement}(d–f). In all cases, $2\theta_0(t)$ increases roughly linearly with time, with a growth rate that decreases as $r$ increases. From linear fits (red dashed lines), the angular spreading rates are estimated to be $1.6^{\circ}\!/\mathrm{h}$ for $r = 200 \ \mathrm{mm}$, $0.9^{\circ}\!/\mathrm{h}$ for $r = 250 \ \mathrm{mm}$, and $0.7^{\circ}\!/\mathrm{h}$ for $r = 300 \ \mathrm{mm}$. 

This decrease in spreading rate with increasing radius reflects the exponential increase of the stationary ice height with $r$ as well as the gaussian distribution of the rivulet residence time with respect to $\theta$: at larger distances from the source, more rivulet passages are required to reach a given thickness, which slows down the growth on the edges of the ice structure. After $18 \mathrm{h}$, Figure~\ref{Exemple epatement}(d-f) suggest that none of the three profiles has reached a quasi-stationary angular width. At that time, the plateau widths are $2\theta_0 \in \{30^\circ, 20^\circ, 16^\circ\}$ for $r \in \{200, 250, 300\} \ \mathrm{mm}$, respectively. Since the opening angle of the lateral envelope in this experiment is approximately $31^\circ$ (as observed in the final height map of Figure~\ref{Macrostructure et differents etats finaux}(a)) this indicates that the section at $r = 200 \ \mathrm{mm}$ has nearly converged to a rectangular $\theta$-invariant profile. If this value is taken as the fully developed stationary width, extrapolating the observed growth rates suggests that the section at $r = 250 \ \mathrm{mm}$ would require about $30 \mathrm{h}$ to converge angularly, and the one at $r = 300 \ \mathrm{mm}$ nearly $41 \mathrm{h}$. These estimates show that after $18 \mathrm{h}$, the ice structures have only partially converged in the angular direction. Overall, the domain that has reached convergence after $18 \mathrm{h}$ is therefore restricted to the region where $r < r_\mathrm{18 h}$ and $\lvert \theta \rvert < \theta_0$.

\section{Conclusion}

In this paper we have investigated the solidification of a water rivulet flowing over a cold inclined substrate, and shown how this configuration generates three-dimensional ice structures over a period of 18 hours. Using a controlled hydraulic and thermal setup, combined with spatiotemporal phase-shifting profilometry and infrared thermography, we analyzed both the temporal evolution and the final morphology of these structures. The dynamics of an experiment typically follow a reproducible sequence: (i) the rivulet first forms a straight ice ridge on which it remains stable; (ii) the rivulet then destabilizes on this ridge, leading to lateral excursions of the rivulet and a rapid transverse expansion of the ice block; and (iii) as the rivulet oscillates and creates new ice branches, the block thickens, gaps between branches gradually fill, and the overall surface becomes smoother. Across all tested conditions, the final ice structure consists of an upstream triangular lateral envelope, transitioning into a downstream zone of constant width once the edges of the substrate are reached (which is generally the case).

We have quantified how the rivulet explores the ice surface using infrared measurements. The total residence time of the flow at the substrate follows a self-similar Gaussian distribution in the azimuthal angle $\theta$, characterized by a standard deviation that is independent of the radial distance $r$. As a consequence, the central region of the structure is visited much more frequently than its lateral edges. By analyzing the angularly averaged height $h_\mathrm{moy}(r,t)$, we have shown that, for each experiment, there exists a convergence radius $r_\mathrm{18 h}$ inside which the height profile has essentially reached a stationary state by the end of the $18 \mathrm{h}$ experiment, while the ice continues to thicken downstream in the region $r > r_\mathrm{18 h}$.

On the converged domain $r < r_\mathrm{18 h}$, we have derived and tested a two dimensional theoretical model for the stationary height field. Extending the framework of \citet{Huerre_2021} to a rivulet flowing over a slowly varying ice slope, we coupled a lubrication-based hydrodynamic description of the rivulet to the heat equations in both liquid and solid phases, linked by a Stefan condition at the interface. In the large Péclet number limit relevant to our experiments, this model predicts an exponential increase of the stationary ice height along the flow direction, characterized by an initial height $h_0$ and a characteristic length $r_c$. Although this behavior was already identified by \citet{Huerre_2021}, the present model demonstrates that this prediction can be generalized to substrates with a slowly varying slope. Moreover, incorporating the three-dimensional geometry of the rivulet refines the relationship between the ice height field and the hydrodynamic parameters of the flow. Exponential fits of the experimental centerline profiles at $t = 18 \mathrm{h}$ indeed show excellent agreement with the model over the converged region. The dependence of the characteristic length with the flow rate put forward a change of slope around $Q \simeq 70 \ \mathrm{mL.min^{-1}}$, marking the transition from a straight-rivulet regime, consistent with the assumptions of the model, to a meandering regime in which a different rivulet geometry must be taken into account. We have also examined the angular dependence of the ice height field. Even when the height has converged along the centerline, it may not have converged near the lateral edges of the heat exchanger, precisely because the rivulet residence time follows a centered Gaussian distribution with the azimuthal angle $\theta$. On this basis, we have proposed a scenario for the evolution of an angular cross-section at fixed radius $r$. The initially narrow ice ridge first spreads laterally; then, at the centerline, the height converges towards the stationary value predicted by the model; finally, a stationary plateau region expands gradually from the centerline, until the entire cross-section eventually becomes $\theta$-invariant. The rate at which this plateau widens decreases with $r$ and with the value of the stationary height: from the theoretical prediction of the stationary profile, we deduce that lower flow rates or higher reduced temperatures lead to taller structures that require more time to converge towards a $\theta$-invariant cross-section. As a result, after $18 \mathrm{h}$, only a restricted region in $(r,\theta)$ space has reached a stationary height, while elsewhere the ice continues to grow after that period of time.

Beyond the specific experimental configuration explored here, our results illustrate how a capillary flow, when coupled to phase change, can generate complex three-dimensional ice morphologies. The experimental approach and theoretical framework developed in this study may be extended in several directions. A first perspective is to investigate whether similar mechanisms govern the formation of ice draperies occurring in underground caves, where a rivulet or a thin film flows at low flow rate over vertical or overhanging cold surfaces. Comparing the present model with such digitized morphologies would help assess its generality. A second perspective is to study the injection of a rivulet directly over an ice surface, thereby producing melting rather than solidification. Such a configuration would bring the system closer to the situation encountered on natural glaciers, where supraglacial streams flow over the ice surface and progressively reshape it.\\

\noindent \textbf{Funding.} This work was supported by Agence de l'Innovation de D\'efense (AID) - via Centre Interdisciplinaire d'Etudes pour la D\'efense et la S\'ecurit\'e (CIEDS) - (project 2021 - ICING).\\

\noindent \textbf{Competing Interests.} The authors report no conflict of interest.\\

\noindent \textbf{Data availability statement.} The data that support the findings of this study are available from the corresponding author, upon reasonable request.\\

\noindent \textbf{Author ORCIDs.} W. Sarlin, https://orcid.org/0000-0002-2668-2279; C. Josserand, https://orcid.org/0000-0003-1429-4209; T. Séon, https://orcid.org/0000-0001-6728-6072; A. Huerre, https://orcid.org/0000-0003-4702-5128.

\appendix

\section{Aperture of the ice structure}\label{Appendix angle de cone}

As one can observe through the different ice final states shown in Figure~\ref{Macrostructure et differents etats finaux}, all structures possess the same type of triangular envelope after 18h of experiment, with an opening half-angle depending on the experimental conditions. Due to the finite width of the heat exchanger, this triangular envelope usually intersects with the edges of the substrate, leading to a region of constant width at the bottom of the exchanger.

\begin{figure}
    \centering
    \includegraphics[width=\linewidth]{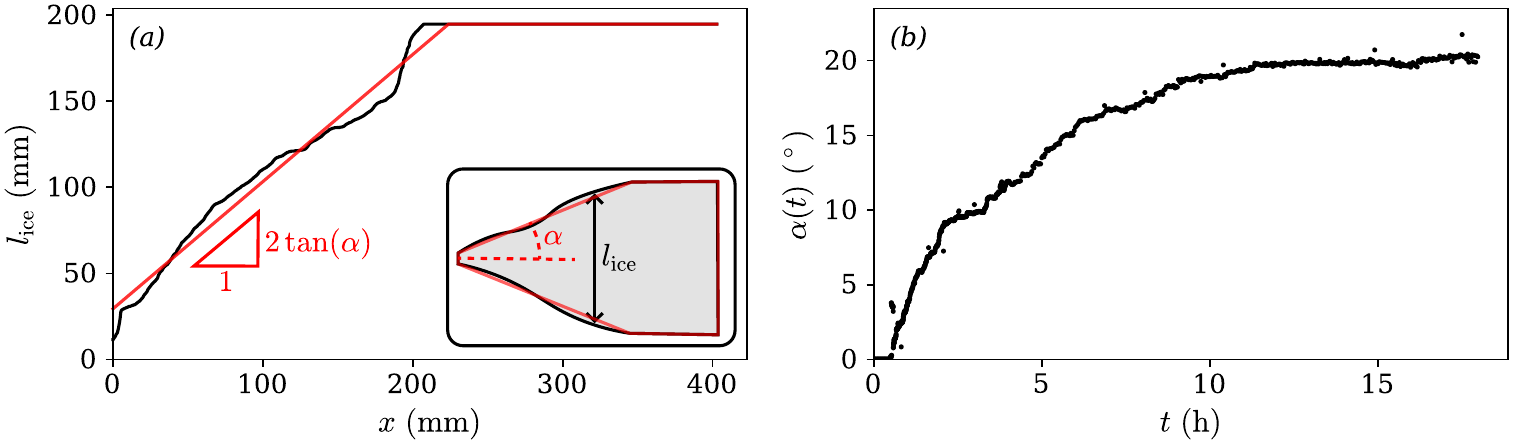}
    \caption{Determination of the opening half-angle of the ice structure’s lateral envelope from the evolution of its width $l_\mathrm{ice}$ during a typical experiment (under the conditions $\alpha_s=50.3^\circ$, $Q=31 \ \mathrm{mL.min^{-1}}$ and $\overline{T}=0.72$). The inset in graph (a) shows a top view of the ice structure and the variation of $l_\mathrm{ice}$ with the cartesian coordinate $x$, obtained from a profilometric reconstruction after $18 \mathrm h$. The curve in graph (a) is fitted with a piecewise linear function (linear then constant), representing an idealized symmetric triangular envelope that ends up intersecting the substrate's edges. From this fit, an opening half-angle $\alpha(t)$ is extracted. Graph (b) shows the temporal evolution of this angle, which converges after about $10 \ \rm h$ to a steady value of $20.1^\circ$.}
    \label{Convergence lateral envelope}
\end{figure}

Before investigating how the aperture depends on the control parameters, we first examine its temporal evolution during a given experiment. To this end, we monitor the lateral spreading of the ice structure through the evolution of its width $l_\mathrm{ice}$ with regard to the distance $x$. This analysis is done in Figure~\ref{Convergence lateral envelope} for an experiment under the conditions $\alpha_s=50.3^\circ$, $Q=31 \ \mathrm{mL.min^{-1}}$ and $\overline{T}=0.72$. For a given time $t$ of the experiment, one can extrapolate an equivalent half-angle $\alpha(t)$ by fitting $l_\mathrm{ice}$ with an affine curve before the structure reaches the exchanger's edges. As an example, this fitting method is shown in Figure~\ref{Convergence lateral envelope}(a) at $t=18 \mathrm h$. One can then extract the temporal evolution of $\alpha(t)$ as depicted in Figure~\ref{Convergence lateral envelope}(b). As the ice structure spreads gradually due to the oscillations of the rivulet, one can observe the increase of $\alpha$ until it reaches a final stationary value $\alpha_\mathrm{fin}=20.1^\circ$ after around 10h. Therefore the lateral envelope of the ice structure has converged towards a final envelope characterized by its half-angle $\alpha_\mathrm{fin}$ long before the end of the experiment. This behavior is consistently observed across all experiments: while the convergence timescale varies with the operating conditions, it always remains shorter than the total duration of the experiments.

\begin{figure}
    \centering
    \includegraphics[width=\linewidth]{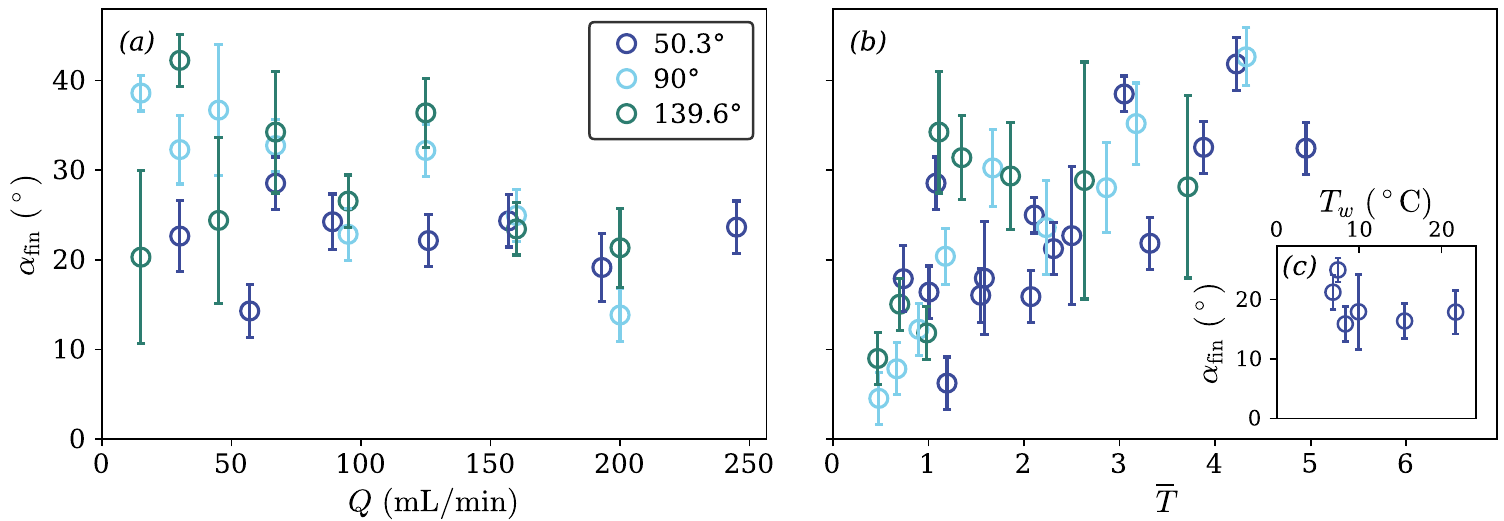}
    \caption{Evolution of the final equivalent half-angle $\alpha_{\rm{fin}}$ of the ice envelope, measured at the end of the experiments ($t=18 \mathrm{h}$), as a function of (a) the flow rate $Q$ at constant temperatures ($T_{\rm{in}}=20^\circ \rm C$ and $T_s=-22^\circ \rm C$), (b) the reduced temperature $\overline{T}$ at constant flow rate ($Q=67 \ \rm{mL.min^{-1}}$), and (c) the injection temperature $T_\mathrm{in}$ at constant flow rate and substrate temperature ($Q=67 \ \rm{mL.min^{-1}}$ and $T_s=-16 \pm 1.2 \ ^\circ \rm C$). For each plot, three experimental series are shown, corresponding to substrate inclinations $\alpha_s \in \{50.3^\circ, \ 90^\circ, \ 139.6^\circ\}$. The half-angles are extracted independently on the left and right sides of the substrate, and their mean defines $\alpha_\mathrm{fin}$. The uncertainty is estimated using a Student distribution that incorporates both the structural asymmetry and a systematic measurement error of $3^\circ$ on each half-angle.}
    \label{Angle avec parametres de controle}
\end{figure}

Based on this result, we can now examine how the control parameters affect the final angle $\alpha_\mathrm{fin}$. The results are presented in Figure~\ref{Angle avec parametres de controle}: each data point corresponds to an individual experiment and represents one of three substrate inclinations, $\alpha_s \in \{50.3^\circ, \ 90^\circ, \ 139.6^\circ\}$, each indicated by a distinct color. As shown in Figure~\ref{Angle avec parametres de controle}(a), the equivalent half-angle $\alpha_\mathrm{fin}$ remains nearly constant with respect to the flow rate $Q$, around $\alpha_\mathrm{fin} \sim 25^\circ$, regardless of the substrate inclination, given the significant uncertainty associated with the measurement method. Experiments conducted at low flow rates exhibit greater variability in the determination of $\alpha_\mathrm{fin}$, with values reaching up to $40^\circ$. When examining each inclination separately, only the vertical configuration ($\alpha_s = 90^\circ$) shows a slight decrease in $\alpha_\mathrm{fin}$. For the other two inclinations, the half-angle fluctuates around an average value without any clear monotonic trend. Overall, these observations suggest that the lateral envelope of the ice structure is not primarily governed by hydrodynamic effects.

To further investigate the mechanism underlying the envelope formation, Figure~\ref{Angle avec parametres de controle}(b) shows the variation of $\alpha_\mathrm{fin}$ with the reduced temperature $\overline{T}$. Unlike the previous plot, this curve exhibits a clear monotonic trend for all substrate inclinations: $\alpha_\mathrm{fin}$ increases with $\overline{T}$. This behavior suggests that higher reduced temperatures allow the rivulet to spread more laterally over the substrate, resulting in a wider envelope. The lateral envelope thus appears mainly governed by thermal effects through phase-change dynamics. However, since $\overline{T}$ depends on both $T_\mathrm{in}$ and $T_s$, their individual contributions cannot be decoupled from Figure~\ref{Angle avec parametres de controle}(b) alone. To address this, we selected six data points already present in the main plot, corresponding to experiments performed at fixed $\alpha_s = 50.3^\circ$ and $Q = 67 \ \mathrm{mL,min^{-1}}$, with nearly constant substrate temperatures $T_s \in [-17.2,-15.6] \, ^\circ\mathrm{C}$, while varying the injection temperature over a wide range $T_\mathrm{in} \in [6.8, 21.7] \, ^\circ\mathrm{C}$. These six experiments are highlighted separately in Figure~\ref{Angle avec parametres de controle}(c). Despite the large variation in $T_\mathrm{in}$, $\alpha_\mathrm{fin}$ remains mostly unchanged, clustering around $19^\circ$. This demonstrates that the injection temperature has little influence on the envelope aperture, which is therefore mainly controlled by the substrate temperature $T_s$.

Overall, the lateral envelope of the ice structure is thus primarily controlled by thermal effects, and more specifically by the substrate temperature $T_s$. Indeed this parameter is known to have, in particular, a strong influence on the apparent contact angle of water on ice \citep{Sarlin_2025}: the larger the undercooling (temperature difference relative to the melting temperature), the greater the apparent angle of water on its substrate. One can expect this change of apparent contact angle to change the overall rivulet dynamics, such as the lateral movement of a rivulet in a dynamic meandering regime. As a consequence, this can impact the angular span in which the rivulet will oscillate laterally, and thus impact the aperture of the ice structure. This provides a plausible explanation for the observed increase of the lateral half-angle $\alpha_\mathrm{fin}$ with decreasing $T_s$, as shown in Figure~\ref{Angle avec parametres de controle}(b). A detailed investigation of contact lines dynamics in the presence of solidification lies beyond the scope of the present study, but would clearly merit a dedicated analysis in the future.

\section{Derivation of the heat equations in curvilinear coordinates}\label{Appendix detail obtention equation theorique hauteur glace}

To derive Equation~\eqref{eq hauteur glace theorique}, we need to rewrite the advection–diffusion equation (Equation~\ref{eq chaleur deux phases}) in a curvilinear coordinate system, expanded at first order in $\varepsilon = 1/\mathrm{Pe} \ll 1$. To this end, the Laplacian can be expressed following the methodology detailed by \citet{Haggstrom_2018}:

\begin{equation}
    \nabla^2T_w = \frac{1}{h_s h_n}[ \frac{\partial}{\partial s}(\frac{h_n}{h_s}\frac{\partial T_w}{\partial s})+\frac{\partial}{\partial n}(\frac{h_s}{h_n}\frac{\partial T_w}{\partial n})]
\end{equation}

where the metric coefficients $h_s$ and $h_n$ are related to the local curvature $\kappa$ of the interface, defined as $\kappa = \dfrac{d^2 h_i}{d s^2} \,\Bigl(1 + \bigl(\dfrac{d h_i}{d s}\bigr)^2\Bigr)^{-3/2}$, via: 

\begin{equation}
\left\{
\begin{array}{l}    
    \begin{aligned}
        & h_s = 1 - n \kappa \\
        & h_n = 1
    \end{aligned}
\end{array}
\right.
\end{equation}

This leads to the following expression for the Laplacian in curvilinear coordinates:

\begin{equation}
    \nabla^2T_w = \frac{\partial^2 T_w}{\partial n^2} - \frac{\kappa}{1 - n \kappa} \frac{\partial T_w}{\partial n} + \frac{d \kappa}{d s}\frac{n}{(1 - n \kappa)^3}\frac{\partial T_w}{\partial s} + \frac{1}{(1 - n \kappa)^2}\frac{\partial^2 T_w}{\partial s^2}
\end{equation}

Introducing the dimensionless variables $\overline{s} = s/(h_{w,0}Pe_0)$, $\overline{n} = n/h_w(s)$, $\overline{\kappa} = \kappa h_{w,0}$, $\Theta_w = (T_w - T_m)/(T_{in} - T_m)$, $\Theta_i = (T_i - T_s)/(T_m - T_s)$ and $\overline{h_i} = h_i/h_{w,0}$, the heat equation in the liquid becomes:

\begin{align}
\label{eq advection diffusion curviligne no1}
    \frac{\sin(\alpha_s - \alpha(\overline{s}))}{\sin(\alpha_s)} \left(\frac{h_w(\overline{s})}{h_{w,0}}\right)^4 \overline{n}(2 - \overline{n}) \frac{\partial \Theta_w}{\partial \overline{s}} \nonumber &= \frac{\partial^2 \Theta_w}{\partial \overline{n}^2}
    - \frac{h_w(\overline{s})}{h_{w,0}} \frac{\overline{\kappa}}{1 - \overline{n} \, \overline{\kappa}} \frac{\partial \Theta_w}{\partial \overline{n}} \nonumber \\& + \varepsilon \frac{d \overline{\kappa}}{d \overline{s}} \left(\frac{h_w(\overline{s})}{h_{w,0}}\right)^2 \frac{\overline{n}}{(1 - \overline{n} \, \overline{\kappa})^3} \frac{\partial \Theta_w}{\partial \overline{s}}
    \\& +\varepsilon^2 \left(\frac{h_w(\overline{s})}{h_{w,0}}\right)^2 \frac{1}{(1 - \overline{n} \, \overline{\kappa})^2} \frac{\partial^2 \Theta_w}{\partial \overline{s}^2}
\end{align}

It is worth noting that, due to the conservation of the quantity $h_w^4(s)\sin(\alpha_s - \alpha(s))$ derived from Equation~\eqref{eq scaling ruisselet}, the factor $\dfrac{\sin(\alpha_s - \alpha(\overline{s}))}{\sin(\alpha_s)} \left(\dfrac{h_w(\overline{s})}{h_{w,0}}\right)^4$ is identically equal to 1. Since $\dfrac{d h_i}{d s}$ is of order 1 in $\varepsilon$, $\kappa$ is thus of order 2 in $\varepsilon$, and the equation simplifies at first order in $\varepsilon$ to:

\begin{equation}
\overline{n}(2-\overline{n}) \frac{\partial \Theta_w}{\partial \overline{s}} = \frac{\partial^2 \Theta_w}{\partial \overline{n}^2}
\end{equation}

A similar reasoning applied to heat diffusion in the ice yields, at the same order:

\begin{equation}
\frac{\partial^2 \Theta_i}{\partial \overline{n}^2} = 0
\end{equation}

Therefore, at first order in $\varepsilon$, the advection–diffusion equation in the liquid and the diffusion equation in the ice reduce to those obtained by \citet{Huerre_2021}, upon replacing the Cartesian coordinates $(\overline{x},\overline{z})$ with the curvilinear coordinates $(\overline{s},\overline{n})$. The theoretical expression for the ice height (Equation~\eqref{eq hauteur glace theorique}) then directly follows from this model.

\bibliographystyle{jfm}
\bibliography{Solidification, Autres, Profilometry, Meanders, Rivulets, TiO2, Braided_films, Drops, Intro}

\end{document}